\documentstyle [12pt,psfig]{article}
\newcommand{\be}{\begin{equation}}
\newcommand{\ee}{\end{equation}}
\newcommand{\bea}{\begin{eqnarray}}
\newcommand{\eea}{\end{eqnarray}}
\newcommand{\ba}{\begin{array}}
\newcommand{\ea}{\end{array}}

\begin{document}
\baselineskip = 16 pt
 \thispagestyle{empty}
 \title{
\vspace*{-2.5cm}
\begin{flushright}
\vspace{-0.3cm}
{\normalsize CERN-TH.7393/94}\\
\end{flushright}
\vspace{1.5cm}
 Higgs and Supersymmetric Particle Signals\\
 at the Infrared Fixed Point  of the Top Quark Mass
\\
 ~\\}
 \author{ M. Carena  and  C.E.M. Wagner \\
 ~\\
 Theory Division, CERN, CH 1211, Geneva, Switzerland. \\
{}~\\
{}~\\
{}~\\
{}~\\
 }
\date{
\begin{abstract}
We  study the  properties  of the
 Higgs and Supersymmetric
particle spectrum,  associated with the infrared fixed point
 solution of the top quark mass
in the Minimal Supersymmetric Standard Model. We concentrate on
the possible detection of these particles, analysing the
deviations from the Standard Model predictions for the
leptonic and hadronic variables measured at LEP and for  the
$b \rightarrow s \gamma$ decay rate.
We consider the low and moderate $\tan \beta$ regime, imposing the
constraints derived from a proper  radiative $SU(2)_L
 \times U(1)_Y$ symmetry breaking and  we study both, the
cases of universal and non--universal soft supersymmetry breaking
parameters at  high energies. In the first
case, for any given value of  the  top quark mass, the Higgs and
supersymmetric particle spectrum is completely determined as a
function of only two soft
supersymmetry breaking parameters, implying
very definite experimental signatures. In the
case of non--universal mass parameters at $M_{GUT}$,
instead, the strong correlations between the sparticle
masses are relaxed, allowing a richer structure for the precision
data variables. We show, however, that the requirement that the low
energy theory proceeds from a grand unified theory with
a local symmetry group including $SU(5)$ strongly
constrains the set of possible indirect experimental signatures.
As a general feature, whenever a significant deviation from the
Standard Model value of the precision data parameters
is predicted, a light sparticle, visible at LEP2, appears in the model.
\begin{flushleft}
{\normalsize CERN-TH.7393/94}\\
{\normalsize July 1994}\\
\end{flushleft}
\end{abstract}}
\maketitle
\newpage

\section{Introduction}
In the present evidence  of a heavy top quark, it is of interest
to study in greater detail the phenomenological implications of the
 infrared fixed point predictions for the top quark mass.
The low energy fixed point structure of the Renormalization Group (RG)
equation of the top quark Yukawa coupling  is
 associated with   large values of this coupling at the high energy
scale, which, however, remain  in the range of validity of perturbation
theory \cite{IR}.
Within  the Minimal Supersymmetric Standard Model (MSSM)
\cite{IR2}, \cite{Dyn},
the infrared fixed point structure determines the value of the top
quark mass as a function of
 $\tan \beta = v_2/v_1$, the ratio
of the two Higgs vacuum expectation values.
In fact,  for a  range of high energy values of
the top quark Yukawa coupling, such that it
 can reach its perturbative limit                             at some
scale $M_X = 10^{14}-10^{19}$ GeV,  the value of the physical top
quark mass is focused to be $M_t =$ 190--210  GeV $ \sin \beta$,
where the  variation in $M_t$ is mainly due to
 a variation in the value of the strong gauge coupling,
$\alpha_3(M_Z) =$ 0.11--0.13.
There is also a small dependence of the infrared fixed point
prediction on the supersymmetric spectrum, which, however, comes mainly
through the dependence on the spectrum of the running of the
strong gauge coupling. Moreover,
considering the MSSM with unification of gauge couplings at a grand
unification scale $M_{GUT}$ \cite{DGR},
the value of the strong gauge coupling is
determined as a function of the electroweak gauge couplings while its
dependence on the SUSY spectrum can be characterized by a single
effective threshold scale $T_{SUSY}$ \cite{LP}-\cite{CPW}.
Thus, the stronger dependence
of the infrared
fixed point prediction on the SUSY spectrum can be parametrized
through $T_{SUSY}$.  There is also
an independent effect coming from supersymmetric threshold
corrections to the Yukawa coupling, which, for supersymmetric
particle
masses smaller or of the order of 1 TeV, may change the top quark mass
predictions in a few GEV,
but without changing the physical picture \cite{Wright}.

The infrared fixed point structure is independent of the particular
supersymmetry breaking scheme under consideration.
On the contrary,
since the Yukawa couplings -- especially if they are strong -- affect
the running of the mass parameters of the theory, once  the infrared
fixed point structure  is present, it  plays a decisive role in
the resulting (s)particle spectrum of the theory,
its predictive power being of course dependent on the
number of initial free independent soft SUSY breaking parameters.
In addition, to assure a proper breakdown of the electroweak symmetry,
one needs to impose conditions on the low energy
mass parameters appearing in the
scalar potential. Indeed, the condition of a proper radiative
SU(2)$_L$ $\times$ U(1)$_Y$ breaking, together with the top quark
Yukawa coupling infrared fixed point structure
 yields interesting correlations among the free
high energy mass parameters of the theory, which then translate into
 interesting predictions for the
Supersymmetric (SUSY) spectrum \cite{COPW} - \cite{Gun}.
Such correlations depend, however, on the exact soft supersymmetry
breaking scheme. In particular, in the minimal supergravity model,
in which   common  masses for all the scalars and gaugino masses at
the high energy scale are
considered, it follows that, once the value of the top quark mass is
given, the whole spectrum is determined as a
function of two parameters \cite{COPW},\cite{CW1}.
In models in which the universality condition  for the high energy
mass parameters is relaxed, the predictions derived from the
infrared fixed point structure are, instead, weaker. Nevertheless,
the infrared fixed point structure implies always an effective
reduction by two in the number of free parameters of the theory.

The infrared fixed point of the top quark mass is interesting
by itself, due to the many interesting
properties associated with its behaviour.
Moreover,  it has been recently observed in the literature that
the condition of bottom-tau Yukawa coupling unification in minimal
supersymmetric grand unified theories
requires large values of the top
quark Yukawa coupling at the unification scale \cite{LP}-\cite{CPW},
\cite{Ramond}-\cite{BABE}.
Most appealing, in the low and moderate $\tan \beta$ regime, for
values of the gauge couplings compatible with recent predictions
from LEP and for the experimentally allowed values of the bottom  mass,
the conditions of gauge and bottom--tau Yukawa coupling unification
predict values of the top quark mass within 10$\%$ of its infrared
fixed point values \cite{LP},\cite{BCPW}.

In section 2 we concentrate on the infrared fixed  point structure of
the Yukawa couplings. In section 3 we present the evolution of the
mass parameters of the theory in the interesting
region of low values of  $\tan\beta$,
to which we shall restrict ourselves for the present study.
Our analysis considers both the case of universal and non--universal
boundary conditions for the soft supersymmetry beaking scalar mass
parameters at the grand unification scale.
In section 4 we investigate the
theoretical constraints associated with a proper breakdown of the
electroweak symmetry and the requirement of stability of the effective
potential by avoiding possible color breaking minima. Complementing
the above constraints with the properties of the top quark infrared
fixed point structure, we define the allowed low energy mass
parameter space as a function of their high energy values. In section
5 we present the results of the above analysis translated into predictions
for the Higgs and supersymmetric particle spectra. In section 6,
a discusion of the precision data variables to be analysed
in the present work is presented.
The results for the experimental variables as
a function of the supersymmetric spectrum is analysed in section 7.
In section 8 we analyse the correlations
between the different experimental
variables and their phenomenological implications.
We reserve section 9 for our  conclusions.

\section{ Infrared Fixed Point Structure}

In the Minimal Supersymmetric Standard Model, with unification
of gauge couplings at some high energy scale $M_{GUT} \simeq 10^{16}$
GeV,
the infrared fixed point structure of the top quark Yukawa coupling may
be easily analyzed,
in the low and moderate $\tan \beta $ regime (1$\leq \tan \beta <$10),
for which
the effects of the bottom and tau Yukawa couplings are
negligible. Indeed, an exact solution for the
running top quark Yukawa coupling may be
obtained  \cite{Ibanez},\cite{Savoy2} in this regime,

\begin{equation}
Y_t(t) = \frac{ 2 \pi Y_t(0) E(t)}{ 2 \pi + 3 Y_t(0) F(t)} ,
\end{equation}
where $E$ and $F$ are functions of the gauge couplings,
\be
E = (1 + \beta_3 t)^{16/3b_3}
(1 + \beta_2 t)^{3/b_2}
(1 + \beta_1 t)^{13/9b_1},
\;\;\;\;\;\;\;\;\;\;\;\; F= \int_{0}^t E(t') dt',
\label{eq:topYuk}
\ee
$Y_t = h_t^2/4\pi$,
$\beta_i = \alpha_i(0) b_i/4\pi$, $b_i$ is  the
beta function coefficient of the gauge coupling $\alpha_i$ and
$t = 2 \log(M_{GUT}/Q)$.
 For large values of $\tan \beta$, instead, the bottom Yukawa
coupling becomes large and, in general,
 a numerical study of the
 coupled equations for the  Yukawa
couplings becomes necessary even at the one
loop level.

For large values of the top quark Yukawa coupling at high energies,
Eq. (\ref{eq:topYuk}) tends to an infrared fixed point value, which
is independent of the exact boundary conditions at $M_{GUT}$, namely,
\begin{equation}
Y_t^{f (Y_t \gg Y_b)}(t) \simeq \frac{2 \pi E(t)}{3 F(t)}.
\label{eq:IR}
\end{equation}
For values of the grand unification scale $M_{GUT} \simeq 10^{16}$ GeV,
the fixed point value,
Eq. (\ref{eq:IR}), is given by $Y_t^f \simeq
(8/9) \alpha_3(M_Z)$. Since in this case
$F(Q = M_Z) \simeq 300$,
the infrared fixed point solution
is rapidly reached for a wide range of values of $Y_t(0) \simeq
0.1 - 1$.  The fixed point structure for the top quark Yukawa coupling
implies an infrared fixed point for
the running  top quark mass,
$m_t(t) = h_t(t) v_2 = h_t(t) v \sin \beta$, with
$v^2 = v_1^2 + v_2^2$,
\be
  m^{IR}_t(t) = h_f(t) \; v\; \sin\beta = m_t^{IRmax.}(t) \; \sin \beta,
\label{eq:mtIR}
\ee
wher we have neglected the slow running of the Higgs vacuum expectation
value at low energies.
For $\alpha_3(M_Z) = 0.11$ -- 0.13,
$m_t^{IRmax.}$ is
approximately given by
\be
 m^{IRmax.}_t(M_t) \simeq 196 GeV  [1+ 2( \alpha_3(M_Z) - 0.12) ].
\label{eq:mtIRmax}
\ee
One should remember that there is a significant quantitative difference
between the running top quark mass,
and the physical
top quark mass $M_t$, defined as the location of the pole in its two
point function. The main source of this difference comes from the
QCD corrections, which at  the one loop level are given by
\be
M_t = m_t(M_t) \left( 1 + 4 \alpha_3(M_t) /3 \pi  \right).
\label{eq:Mt}
\ee
A numerical
two loop RG analysis, shows the stability of
the infrared fixed point under higher order loop
contributions \cite{BABE},\cite{CPW}.

A similar exact
analytical study can be done for the large $\tan \beta$ regime,
when the bottom and top Yukawa couplings are equal at the unification
scale, by
neglecting in a first approximation the effects of the
tau Yukawa coupling  and identifying the right-bottom and right-top
hypercharges. The approximate solution for
$Y= Y_t \simeq Y_b$ reads \cite{wefour},
\be
Y(t) = \frac{ 4 \pi Y(0) E(t)}{ 4 \pi + 7 Y(0) F(t)}
\ee
Then, if the Yukawa coupling is large at the grand unification scale,
at energies of the order of the top quark mass it will develop an
infrared fixed point value approximately given by \cite{CPW},
\cite{wefour},
\be
Y_t^{f\;(Y_t=Y_b)}(t) \simeq \frac{4 \pi E(t)}{7 F(t)}
\simeq  \frac{6}{7} Y_t^{f(Y_t \gg Y_b)}(t).
\ee
An approximate expression for the fixed point solution may be found
also for values of the bottom Yukawa coupling different from the
top quark one \cite{CW}.

In general, in the large $\tan\beta$ region
the bottom quark
Yukawa coupling becomes strong and plays an important role in the
RG analysis \cite{somp}. The are also possible
large radiative corrections to the bottom
mass coming from  loops of supersymmetric particles, which are strongly
dependent on the  particular spectrum and are extremely
important in  the
analysis, if unification of bottom and tau
Yukawa couplings is  to  be considered \cite{Hall}--\cite{wefour}.
Moreover,
 in some of the minimal models of grand unification,
large $\tan \beta$ values are in conflict with proton
decay constraints \cite{AN}.
In the special case of  tau-bottom-top Yukawa coupling unification the
infrared fixed point solution for the top quark mass is not achievable
unless a relaxation in the high energy boundary conditions of the
mass parameters of the theory is arranged, and it is necessarily
associated with a heavy supersymmetric spectrum \cite{nonunl}.
In the following we shall concentrate on the low and moderate
$\tan \beta$ region.

\section{Evolution of the Mass Parameters}

In this work  we
 shall consider soft supersymmetry breaking mass terms for all
the scalars and gauginos of the theory, as well as
trilinear and bilinear couplings
 $A_i$ (with i= leptons, up quarks and down quarks)  and $B$ in the
full scalar potential, which are proportional to the
trilinear and bilinear terms appearing in the
superpotential.
In the framework of minimal supergravity the
soft supersymmetry breaking parameters are universal
at the grand unification
scale. This implies
 common soft supersymmetry
breaking mass terms $m_0$ and $M_{1/2}$ for the scalar and
gaugino sectors of the theory, respectively, and a common
value $A_0$ ($B_0$) for all trilinear (bilinear)
couplings $A_i$ ($B$).
In addition, the supersymmetric Higgs mass parameter $\mu$ appearing
in the superpotential takes a value $\mu_0$ at the grand unification
scale $M_{GUT}$.
In the present work we shall consider a more general case, in which the
condition of universality of the soft supersymmetry breaking scalar mass
parameters is relaxed. We shall, however, asume that
SU(5) is a subgroup of the grand unification symmetry group and, hence,
we shall keep the relations between the soft supersymmetry breaking
mass parameters  that  preserve the SU(5) symmetry. This implies
a common gaugino soft supersymmetry breaking mass parameter, common
values for the soft supersymmetry breaking parameters of the right
and left handed scalar top quarks, but free, independent values for
the two Higgs mass parameters at $M_{GUT}$. The relevance of
non--universal soft supersymmetry breaking parameters for the
spectrum of the theory in the low $\tan\beta$ regime
has been recently emphasized in several works \cite{nonun}.
For definiteness, we
shall identify all squark and slepton mass parameters with the ones
of the stop quark ones. This requirement has little influence in our
analyisis, which mainly depends on the Higgs, chargino and stop
spectra.

Knowing the values of
the mass parameters at the unification scale, their low energy
values may be specified by their renormalization group evolution
\cite{Ibanez}-\cite{BG},
which contains also  a dependence on the gauge and Yukawa
 couplings.
In particular, in the low and moderate $\tan \beta$ regime, in which
the effects of the bottom and tau Yukawa couplings are negligible, it
is possible to determine  the evolution of the soft supersymmetry
breaking mass parameters of the model as a
function of  their high energy boundary conditions
and the value of the top quark Yukawa coupling at $M_{GUT}$, $Y_t(0)$.
Indeed, using Eq. (\ref{eq:IR})
and renaming $Y_t^{f(Y_t \gg Y_b)} = Y_f(t)$, it follows,
\begin{equation}
 \frac{6 Y_t(0) F(t)}{4 \pi} =
\frac{Y_t(t)/ Y_f(t)}{ 1 - Y_t(t)/ Y_f(t)}\;,
\end{equation}
with $Y_t/Y_f = h_t^2/ h^2_f$
the ratio of Yukawa couplings squared  at low energies.
The above equation permits to express the boundary condition of the
top quark Yukawa coupling as a function of the
gauge couplings (through F) and the ratio  $Y_t / Y_f$
\cite{Ibanez}-\cite{BG} giving
 definite predictions for the low energy mass parameters of the model
in the limit $h_t \rightarrow h_f$ \cite{COPW}.

Thus, considering the limit of small $\tan\beta$, $\tan\beta < 10$,
the following approximate
analytical solutions  are obtained for the case of
non--universal parameters at $M_{GUT}$,
\bea
m_L^2 & =&  m_L^2(0) + 0.52  M_{1/2}^2
\;\;\;\;\;\;\;\;\;\;\;\;\;\;\;\;\;\;\;\;\;\;
m_E^2  =  m_E^2(0) + 0.15  M_{1/2}^2
\nonumber \\
\nonumber \\
m_{Q(1,2)}^2 & =&  m_{Q(1,2)}^2(0) + 7.2  M_{1/2}^2
\;\;\;\;\;\;\;\;\;\;\;\;\;\;\;\;
m_{U(1,2)}^2  \simeq  m_{U(1,2)}^2(0) + 6.7  M_{1/2}^2
\nonumber \\
\nonumber \\
 m_D^2 & \simeq & m_D^2(0) + 6.7  M_{1/2}^2
\nonumber \\
\nonumber \\
m_Q^2 &=& 7.2 M_{1/2}^2 + m_Q^2(0) + \frac{\Delta m^2}{3}
\nonumber \\
m_U^2 &=& 6.7 M_{1/2}^2 + m_U^2(0) + 2 \frac{\Delta m^2}{3}
\label{eq:todas}
\eea
where E, D and U are the right handed leptons, down-squarks and
up-squarks, respectively,  L  and Q =  (T B)$^T$ are the  lepton and
top-bottom left handed doublets and $m_{\eta}^2$,
with $\eta=E,D,U,L,Q$
are the corresponding soft supersymmetry breaking
mass parameters.
The subindices (1,2) are to distinguish
the first and second generations from the third one, whose
mass parameters receive the
top quark Yukawa coupling contribution to their renormalization
group evolution, singled
out in the $\Delta m^2$ term,
\begin{eqnarray}
\Delta m^2 &  = & - \frac{ \left(m^2_{H_2}(0) + m_U^2(0)
+ m_Q^2(0)\right) }{2}
 \frac{Y_t}{Y_f} + 2.3 A_0 M_{1/2}
\frac{Y_t}{Y_f} \left( 1 - \frac{Y_t}{Y_f} \right)
\nonumber\\
& - &
\frac{A_0^2}{2} \frac{Y_t}{Y_f} \left( 1 - \frac{Y_t}{Y_f} \right)
+ M_{1/2}^2
\left[
- 7 \frac{Y_t}{Y_f} + 3
\left(
\frac{Y_t}{Y_f} \right)^2 \right] \; .
\label{eq:dm}
\end{eqnarray}
For the Higgs sector, the mass parameters involved are
\be
m_{H_1}^2  =  m_{H_1}^2(0) + 0.52  M_{1/2}^2  \;\;\;\;\;
and \;\;\;\;\;\;
m_{H_2}^2 = m_{H_2}^2(0) + 0.52 M_{1/2}^2  + \Delta m^2\; ,
\label{eq:m12}
\end{equation}
which are the
soft supersymmetry breaking parts of the  mass parameters $m_1^2$ and
$m_2^2$ appearing in the  Higgs scalar potential (see section 4).
Moreover,
 the  renormalization group
evolution for the supersymmetric mass parameter $\mu$
reads,
\begin{equation}
\mu^2 \simeq 2 \mu_0^2 \left( 1 - \frac{Y_t}{Y_f} \right)^{1/2}  \; ,
\label{eq:mu}
\end{equation}
while the running of the soft supersymmetry breaking bilinear
 and trilinear couplings gives,
\begin{equation}
B = B_0 - \frac{A_0}{2} \frac{Y_t}{Y_f} + M_{1/2} \left(1.2
\frac{Y_t}{Y_f} - 0.6 \right).
\label{eq:b0}
\end{equation}
\begin{equation}
A_t = A_0 \left(1 - \frac{Y_t}{Y_f} \right) - M_{1/2} \left(4.2 - 2.1
\frac{Y_t}{Y_f} \right),
\label{eq:a0}
\end{equation}
respectively.
Eq. (\ref{eq:mu}) shows that the RG evolution of
the supersymmetric mass parameter $\mu$, appearing in the
superpotential. Observe that $\mu$ formally vanishes at low energies
in the limit $Y_t \rightarrow Y_f$. However, since $\mu \simeq
\sqrt{2} \mu_0 \; (1 - Y_t/Y_f)^{1/4}$, $\mu$ stays of order $\mu_0$
even for values all values of $Y_t$ within the range of validity
of perturbation theory at high energies, $Y_t(0) \leq 1$
($Y_t/Y_f \leq 0.995$) \cite{COPW}.
 The coefficients characterizing the
 dependence of the mass parameters on the universal gaugino
mass $M_{1/2}$  depend
on the exact value of the strong  gauge
couplings. In the
above, we have taken the values of the coefficients that
are  obtained for $\alpha_3(M_Z) \simeq 0.12$.
The above  analytical solutions are    sufficiently accurate for
the purpose of        understanding    the properties of
the mass parameters in the limit $Y_t \rightarrow Y_f$.

\section{Constraints on the Fixed Point Solutions}

The solutions for the mass parameters may be strongly constrained
by experimental and theoretical restrictions. The experimental
contraints come  from the   present lower bounds on the supersymmetric
particle masses \cite{Partd}.
Concerning the  theoretical constraints, many of them
impose  bounds on the allowed space for
 the soft supersymmetry breaking parameters in  model
 dependent ways to various degrees. The conditions of stability of
  the effective potential and a proper breaking of the SU(2)$_L$
$\times$ U(1)$_Y$ symmetry  are, instead, basic
necessary requirements, which, complemented with the properties derived
from the infrared fixed point structure, yield robust correlations
among the free parameters of the theory.

\subsection{Radiative electroweak symmetry breaking}

The Higgs potential
 of the Minimal Supersymmetric
Standard Model may be written as \cite{Dyn}, \cite{CSW}-\cite{HH}
\begin{eqnarray}
V_{eff} & = & m_1^2 H_1^{\dagger} H_1 +
m_2^2 H_2^{\dagger} H_2 - m_3^2 (H_1^T i \tau_2 H_2
+ h.c.)
\nonumber\\
& + & \frac{\lambda_1}{2} \left(H_1^{\dagger} H_1 \right)^2
+ \frac{\lambda_2}{2} \left(H_2^{\dagger} H_2 \right)^2
+ \lambda_3 \left(H_1^{\dagger} H_1 \right)
 \left(H_2^{\dagger} H_2 \right)
+ \lambda_4 \left| H_2^{\dagger} i \tau_2 H_1^* \right|^2 ,
\end{eqnarray}
with
  $m_i^2 = \mu^2 + m_{H_i}^2$,  $i = 1,2$, and   $m_3^2 = B |\mu|$ and
where at scales at which the theory is supersymmetric  the
running quartic couplings $\lambda_j$, with $j = 1 - 4$,
must satisfy the following conditions:
\begin{equation}
\lambda_1 = \lambda_2 = \frac{ g_1^2 + g_2^2}{4} = \frac{M_Z^2}{2\;v^2}
,\;\;\;\;\;
\lambda_3 = \frac{g_2^2 - g_1^2}{4},\;\;\;\;\;
\lambda_4 = - \frac{g_2^2}{2} = \frac{M_W^2}{v^2}.
\end{equation}
Hence, in  order to obtain the low energy values of the quartic
couplings, they must  be evolved using
the appropriate
renormalization group equations, as was explained in
Refs. \cite{CSW}-\cite{Chankowski}.
  The mass parameters $m_i^2$, with $i = 1$-$3$ must
also be evolved in a consistent way  and their RG equations may be
found in
the literature \cite{Ibanez}-\cite{Savoy2},\cite{Inoue},\cite{OP}.
  The minimization conditions
$\partial V/ \partial H_i |_{<H_i>=v_i} =0$, which are necessary
to impose the proper breakdown of the electroweak symmetry,
read
\begin{equation}
\sin(2\beta) = \frac{ 2  m_3^2  }{m_A^2}
\label{eq:s2b}
\end{equation}
\begin{equation}
\tan^2\beta = \frac{m_1^2 + \lambda_2 v^2 +
\left(\lambda_1
 - \lambda_2 \right) v_1^2}{m_2^2 + \lambda_2 v^2},
\label{eq:tb}
\end{equation}
where
$m_A$ is the CP-odd Higgs
mass,
\begin{equation}
m_A^2 = m_1^2 + m_2^2 + \lambda_1 v_1^2 +
\lambda_2 v_2^2 + \left( \lambda_3 + \lambda_4 \right) v^2 .
\end{equation}

Considering the case of negligible stop mixing, and in the low
$\tan\beta$ regime, the radiative corrections to the quartic
couplings $\lambda_i$, with $i = 1, 3$ are small, while
$\Delta \lambda_2 = (3/ 8 \pi^2) h_t^4 \ln (m_{\tilde{t}}^2/m_t^2)$.
In this case,
the minimization condition  Eq. (\ref{eq:tb}),
can be rewritten as \cite{CSW}:
\begin{equation}
\tan^2\beta = \frac{m_1^2 + M^2_Z/2}{m_2^2 + M^2_Z/2 +
\Delta \lambda_2 v_2^2}.
\label{eq:tb2}
\end{equation}
Considering the minimization condition,
Eq. (\ref{eq:tb}) and
the approximate analytical expressions for
the mass parameters $m_i$, Eq. (\ref{eq:m12}),
the supersymmetric mass parameter $\mu$ is determined as a function of
the other free  parameters of the theory,
\bea
\mu^2 &=&  \frac{1}{\tan^2 \beta -1} \left(
m_{H_1}^2 - m_{H_2}^2 \tan^2 \beta -
\Delta \lambda_2 v_2^2 \tan^2 \beta \right)
\nonumber \\
 &=& {\cal{F}}( m_{H_1}(0), m_{H_2}(0), m_Q(0), m_U(0), M_{1/2},
 A_0, \tan \beta, Y_t /Y_f)    .
\label{eq:calF}
\eea
A somewhat more complicated expression is obtained for the case of
mixing in the stop sector \cite{HH}.
The other minimization condition, Eq. (\ref{eq:s2b})
also puts restrictions
on the soft  supersymmetry breaking
 parameters. It determines the value of the
parameter $\delta = B_0 - A_0$ as a function of the other parameters
of the theory \cite{COPW}. However, as we shall show below,
at the fixed point solution the mass parameter
$\delta$ is not directly related to the range of possible mass values
of the Higgs and supersymmetric particles.

\subsection{Properties of the Fixed Point Solution.}

The ratio of the top quark Yukawa coupling to its infrared fixed
point value may be expressed as a function of the top quark mass and the
angle $\beta$,
\be
\frac{Y_t}{Y_f} = \left( \frac{m_t}{m_t^{IRmax}} \right)^2
\frac{1}{\sin^2 \beta},
\label{eq:Y}
\ee
where the exact value of $m_t^{IRmax.}$,
Eq. (\ref{eq:mtIRmax}), depends
on the value of the strong gauge coupling  considered, and
 for the experimentally allowed range it varies approximately between
190--200 GeV.
Depending on the precise value of the running
top quark mass $m_t$ and $\tan \beta$,
the above equation gives a measure of the proximity to the infrared
fixed point solution.
In the limit $Y_t \rightarrow Y_f$,
 the strong
correlation between the top quark mass and the value of $\tan\beta$,
Eq.(\ref{eq:mtIR}),
 allows to reduce by one   the number of
free parameters of the theory.

 Moreover, at the infrared fixed point,
the  expressions for the low energy  parameters, Eqs.
(\ref{eq:todas})-(\ref{eq:a0}),
 show that the term $\Delta m^2$ and hence the
 mass parameters  $m_{H_2}^2$, $m_Q^2$ and
$m_U^2$ become  very weakly dependent on the supersymmetry
breaking parameter $A_0$. In fact, the dependence on $A_0$ vanishes
in the formal limit $Y_t \rightarrow Y_f$ \cite{COPW}.
The only relevant dependence on $A_0$
enters through the mass parameter $B$.
Therefore, at the infrared fixed point, there
 is an effective reduction in
the number of free independent soft
supersymmetry breaking parameters. In fact, the dependence on
$B_0$ and $A_0$ of the low energy solutions is effectively
replaced by
a dependence on the parameter $\delta = B_0 - A_0/2$.
Since B is not involved in the RG evolution of the
(s)particle masses and the squark and slepton mixing for sparticles
other than the top squark is very small,
the above implies that at the infrared fixed point
the dependence of the Higgs and supersymmetric
spectrum on the parameter $A_0$ is negligible \cite{COPW}.
Hence, the infrared fixed point structure translates in a net reduction
by two in the number of free parameters which are relevant in
determining the spectrum of the theory.
There is also a very interesting behaviour  of the low energy
mass parameter combination
\begin{equation}
M_{UQ}^2 =  m_Q^2 + m_U^2 + m_{H_2}^2
\end{equation}
at the infrared fixed point. Indeed, the dependence of $M_{UQ}$ on its
high energy boundary condition,
 $M_{UQ}^2(0) =  m_Q^2(0) + m_U^2(0) + m_{H_2}^2(0)$
 vanishes in the
formal limit $Y_t \rightarrow Y_f$.
It follows that the infrared fixed point structure of the top
quark Yukawa coupling yields an infrared fixed point for  the soft
supersymmetry breaking  parameter $A_t$ as well as for the
combination $M_{UQ}^2$.

Summarizing,
for a given value of the
physical  top quark mass, the running top quark mass is fixed and then
at the infared fixed point
Eq. (\ref{eq:Y}) fixes  $\sin \beta$. Due to the strong correlation of
the top quark mass with $\tan\beta$ and
the independence of the spectrum on the parameter
$A_0$,  for a given top quark mass the Higgs and
supersymmetric particle spectrum is completely determined as a function
of only the
high energy boundary conditions for the scalar and gaugino mass
mass parameters.
It is then possible to perform a scanning of all the possible values
for $m_Q(0)$ ($m_Q(0) \equiv m_U(0)$), $m_{H_1}(0)$, $m_{H_2}(0)$
 and $M_{1/2}$,  bounding the squark masses to be, for example,
below 1 TeV, and the whole allowed parameter space for the Higgs and
superparticle masses may be studied.  In the
following we shall study different  boundary conditions for the soft
supersymmetry breaking mass
parameters, concentrating on those which
may yield interesting features for
the low energy spectrum. In particular, we shall also consider
the case in which all soft supersymmetry breaking
scalar masses acquire a common value at the high
energy scale, which gives an extremely predictive framework
with only two parameters determining the whole Higgs and supersymmetric
spectrum.

\subsection{Color breaking minima}

There are several conditions which need to be fulfilled to
ensure the stability of the electroweak symmetry breaking
vacuum. In particular, one should check that no charge or color
breaking minima are induced at low energies. A well
known condition for the  absence
of color breaking minima is given by the relation \cite{ILEK}
\begin{equation}
A_t^2 \leq 3 (M_{UQ}^2) + 3 \mu^2.
\end{equation}
At the fixed point, however, since $A_t \simeq -2.1 M_{1/2}$
and $M_{UQ}^2 \simeq 6 M_{1/2}^2$, this
relation is trivially fulfilled \cite{Nir}, \cite{CW1}.
 For values of $\tan\beta$ close to one, large values of $\mu$ are
induced and  a more appropriate
relation is obtained by looking for possible color breaking
minima in the direction
$\langle H_2 \rangle \simeq \langle H_1 \rangle$ and
$\langle Q \rangle \simeq \langle U \rangle$ \cite{Drees},\cite{Nir}.
The requirement of stability of the physically acceptable
vacuum implies  the following sufficient condition
\cite{CW1},
\begin{equation}
\left( A_t - \mu \right)^2 \leq 2 \left( m_Q^2 + m_U^2 \right)
+ \tilde{m}_{12}^2
\label{eq:cond1}
\end{equation}
where $\tilde{m}_{12}^2 =
\left( m_1^2 + m_2^2 \right)  (\tan\beta - 1)^2 /
(\tan\beta^2 + 1 ) $.

If Eq.(\ref{eq:cond1}) is not fulfilled, a second sufficient
condition is given by
\begin{equation}
\left[ \left( A_t - \mu \right)^2 - 2 \left( m_Q^2 + m_U^2 \right)
- \tilde{m}_{12}^2 \right]^2 \leq 8 \left( m_Q^2 + m_U^2 \right)
\tilde{m}_{12}^2.
\label{eq:cond2}
\end{equation}
The above relations, Eqs. (\ref{eq:cond1}) and (\ref{eq:cond2})
are sufficient conditions since they assure that a color breaking
minima lower than the trivial minima does not develop in the theory.
If the above conditions are violated, a necessary condition to
avoid the existence of a color breaking minima lower than
the physically acceptable one is given by
\begin{equation}
V_{col} \geq V_{ph}
\label{eq:cond3}
\end{equation}
with
\begin{equation}
V_{col} = \frac{(A_t - \mu)^2 \alpha_{min}^2}{ h_t^2 (2 \alpha_{min}^2
+ 1 )^3 } \left[ ( m_Q^2 + m_U^2) - 2 \tilde{m}_{12}^2 \alpha_{min}^4
\right],
\end{equation}
\begin{equation}
V_{ph} = - \frac{M_Z^4}{2 ( g_1^2 + g_2^2)} \cos^2 (2\beta),
\end{equation}
and
\begin{equation}
\alpha_{min}^2 = \left[(A_t - \mu)^2 - 2 (m_Q^2 + m_U^2)
- \tilde{m}_{12}^2 \right]/(4 \tilde{m}_{12}^2).
\end{equation}

For some of the non-universal conditions one may consider, the
right handed stop supersymmetry breaking mass parameter
$m_U^2$ can get negative values. In this case a color
breaking minimum may develop in the direction $\langle U \rangle
\neq 0$. The value of the tree level potential at this
minimum would be
\begin{equation}
V_{U} = - \frac{9}{8 g_1^2} m_U^4, \;\;\;\;\;\;\;\;\;\;
(m_U^2 < 0)
\label{eq:minimu}
\end{equation}
which should be higher than $V_{ph}$ in order to avoid an unacceptable
vacuum state. For low values of $\tan\beta \leq 1.5$,
the range of parameters
leading to a negative value of $m_U^2$, are automatically excluded
either because they lead to values of the lightest
CP-even Higgs mass which are experimentally excluded (particularly for
$\mu < 0$) or because they lead to tachyons in the stop sector or are
in conflict with the absence of color breaking minima in the
other directions analysed before,
Eqs. (\ref{eq:cond1})--(\ref{eq:cond3}).
For larger values of $\tan\beta$
and negative values of $\mu$, for which the mixing in the stop
sector is small, the requirement $V_{U} \geq V_{ph}$, with $V_U$
given in Eq. (\ref{eq:minimu}), becomes, however, relevant.

\section{Higgs and Supersymmetric Particle Spectrum}

As we mentioned before,
 for low values of $\tan\beta$ and for a given
value of the top quark mass, the whole spectrum is determined
as a function of  the  free independent  soft supersymmetry breaking
parameters, $m_Q(0)$
(where $m_Q(0) = m_{\tilde{q}}(0) = m_{\tilde{l}}(0)$),
$m_{H_1}(0)$, $m_{H_2}(0)$
 and $M_{1/2}$.
Summarizing the results for the relevant
low energy mass parameters at the fixed point solution we have:
\begin{eqnarray}
m_{H_2}^2 & \simeq &  \frac{m_{H_2}^2(0)}{2} - m_Q^2(0)
- 3.5 M_{1/2}^2
\;\;\;\;\;\;\;\;\;\;\;\;\;\;
m_{H_1}^2  \simeq  m_{H_1}^2(0) + 0.5 M_{1/2}^2
\nonumber\\
m_Q^2 & \simeq & \frac{2 m_Q^2(0)}{3} - \frac{m_{H_2}^2(0)}{6}
  + 6 M_{1/2}^2
\nonumber \\
m_U^2 & \simeq & \frac{m_Q^2(0)}{3} - \frac{m_{H_2}^2(0)}{3}
+ 4 M_{1/2}^2
\nonumber\\
A_t & \simeq & - 2.1 M_{1/2}
\nonumber\\
\mu^2 & \simeq & \left[
m_{H_1}^2(0) +  \left(\frac{2 m_Q^2(0) - m_{H_2}^2(0)}{2}
\right) \tan^2 \beta
\right.
\nonumber\\
& + & \left.
M_{1/2}^2 \left( 0.5 + 3.5 \tan^2\beta \right) \right]
\frac{1}{ \tan^2\beta - 1 }
\label{eq:massp}
\end{eqnarray}

In the following,
we shall analyze the  contributions of possible light
 supersymmetric particles to the hadronic and leptonic variables
measured at LEP. In fact, concerning indirect searches at LEP, the
existence of light charginos and stops may yield interesting
supersymmetric signals. The Higgs sector is of course very interesting
in itself and it
also plays an important role in deriving constraints
on the soft supersymmetry breaking parameters, which then translate
into restrictions for the stop chargino sectors as well.
The dependence of the properties of the spectrum on the high energy
boundary conditions for the soft supersymmetry breaking parameters is
very important and we shall consider
different interesting possibilities
in a detailed way.  In Table 1 we display the dominant
dependence of the low energy scalar mass parameters on their
high energy values for the three characteristic soft supersymmetry
breaking schemes we shall analyse in this work.
The case of universal soft supersymmetry breaking
parameters, in which all the soft
supersymmetry breaking scalar masses acquire a common value, say $m_0$,
 is the most
predictive one. In such limit, for a given value of the top quark mass,
the whole Higgs and supersymmetric
 particle spectrum is determined as a function of only two parameters,
$m_0$ and the common gaugino mass $M_{1/2}$ \cite{COPW}.

Another interesting case is the one in which the dependence of
$\mu^2$ on the soft supersymmetry breaking parameters of the scalar
fields
vanishes, implying smaller values for the supersymmetric mass
parameter and hence a stronger Higgsino component of the light chargino
than in the universal case. This situation
follows, in a $\tan \beta$ independent way, if $m_{H_1}(0)$ = 0 and
$m_{H_2}^2(0) = 2 m_Q^2(0)$.  As can be seen in Table 1, the parameter
$m_U^2$ can be render small or negative by increasing $m_0$, what
increases the right handed component of the lightest stop with respect
to the one in the case of universal soft supersymmetry breaking
parameters at $M_{GUT}$.  As we shall
discuss below, a larger Higgsino (right stop) component of the lightest
chargino (stop) implies  an increase in the
supersymmetric $Z^0$--$b\bar{b}$ vertex corrections.

A non--universal
condition for the
scalar soft supersymmetry breaking
mass parameters can also yield larger values for the stop mass
parameters. This situation may be achieved if, for example,
we invert the
relations used above
for $m_{H_1}^2(0)$ and $m_{H_2}^2(0)$, that is to say,
$m_{H_1}^2(0) = 2 m_Q^2(0)$ and  $m_{H_2}^2(0) = 0$.
Then, parametrizing the scalar masses as a function of
$m_{H_1}(0)$,
the values of $\mu^2$ and $m_Q^2 + m_U^2$ have the same functional
dependence on $m_{H_1}^2(0)$
as the one as a function of $m_0^2$
in the universal case. However, both $m_Q^2$ and $m_U^2$ increase
with $m_{H_1}^2(0)$, breaking the strong correlation present
in the universal case between the lightest stop and the
gaugino masses \cite{COPW}, \cite{nonun}.
Indeed, even for light charginos, large values
of the lightest stop mass may be obtained in this case by
taking large values for the scalar mass parameters at the grand
unification scale.
{}~\\

\baselineskip = 25pt
\begin{center}
\begin{tabular}{|l|c|c|c|c|}
\hline
   &   &   &   &   \\
Conditions at $M_{GUT}$ &$\;\; m_Q^2 \;\;$  & $\;\; m_U^2 \;\;$
& $\;\; m_{H_2}^2 \;\;$
& $\;\; m_{H_1}^2 \;\;$
\\
   &   &   &   & \\
\hline
   &   &   &   & \\
Universal  $\; m_0^2$
 & {\large $\frac{m_0^2}{2}$}  & 0
&{\large $- \frac{m_0^2}{2}$}  & $m_0^2$
\\
  &  &  &  &
\\
\hline
Case I:  &   &  & &  \\
$m_{H_1}^2(0) = 0 \;$,
$m_{H_2}^2(0) = 2 m_Q^2(0)$
& {\large $\frac{m_0^2}{6}$}
& {\large $- \frac{m_0^2}{6}$} & 0 & 0
\\
$m_{H_2}^2(0) = m_0^2$   &  &  &  &  \\
\hline
Case II:  &   &  &  & \\
$m_{H_2}^2(0) = 0 \;$,
$m_{H_1}^2(0) = 2 m_Q^2(0)$
& {\large $\frac{m_0^2}{3}$}
& {\large $\frac{m_0^2}{6}$}
& {\large $- \frac{m_0^2}{2}$} & $m_0^2$
\\
$m_{H_1}^2(0) = m_0^2$  &  &  &  & \\
\hline
\end{tabular}
\end{center}
{}~\\
\baselineskip = 10pt
{\small
Table 1.  Dominant
dependence of the low energy soft supersymmetry breaking
parameters on their values at the grand unification scale, for
a top quark mass at its infrared fixed point value.}\\
{}~\\
\baselineskip = 16pt
{}~\\
The value of the supersymmetric
mass parameter $\mu$ may be obtained from the other mass
parameters through the condition of a proper radiative electroweak
symmetry breaking, Eqs. (\ref{eq:tb}), (\ref{eq:massp}).
In the following, we shall analyze these three possibilities
considering cases I and II
as characteristic ones for the study of
the possible implications of the deviations from the universal
boundary conditions in the soft supersymmetry breaking parameters
associated with the scalar sector.

\subsection{Stop and Chargino Sectors}

Due to the large values of the  mass parameter
$\mu$ at the infrared fixed point for low values of $\tan\beta$,
there is small mixing in the chargino and
neutralino sectors. Hence, to a good approximation the lightest
chargino mass and the lightest and next to lightest neutralino masses
 are given by $m_{\tilde{\chi}^{\pm}_l} \simeq
m_{\tilde{\chi}^0_2} \simeq
 2 m_{\tilde{\chi}^0_1} \simeq 0.8 M_{1/2}$. This
approximate dependence becomes more accurate when  larger values
of $M_{1/2}$ are considered. For low values of $M_{1/2}$, although
the lightest chargino is still mainly a wino,
it follows that for positive values of $\mu$, slightly larger
values of $M_{1/2}$ are necessary to get a light chargino, with
mass close to their production threshold at the $Z^0$ peak, than
those required for negative
values of $\mu$.

Concerning the restrictions on the parameter space,
more interesting than the chargino sector becomes
the stop sector, for which these large values of $\mu$
may render the physical squared stop mass negative
or too small to be consistent with the present experimental bounds,
which we shall take to be $m_{\tilde{t}} > 45$ GeV.
The stop mass  matrix is given by,
\bea
M^2_{\tilde{t}} = \left[
\begin{tabular}{c c}
   $m_Q^2 + m^2_t + D_{t_L}$ & $ m_t (A_t - \mu/ \tan \beta)$ \\
  $  m_t (A_t - \mu/ \tan \beta)$  & $m_U^2 + m^2_t + D_{t_R}$
\end{tabular} \right]
\label{eq:stopmat}
\eea
where  $D_{t_L} \simeq - 0.35 M_Z^2 |cos 2 \beta|$ and
 $D_{t_R} \simeq - 0.15 M_Z^2 |cos 2 \beta|$ are the D-term contributions
to the left and right handed stops, respectively. The above mass matrix,
after diagonalization, leads to the two stop mass eigenvalues,
 $m_{\tilde{t}_1}$ and  $m_{\tilde{t}_2}$.
At the infrared fixed point,
the values of the  parameters involved in the mass matrix  are given in
Eq. (\ref{eq:massp}). For
values of  $\tan \beta$ close to one, the off-diagonal term
contribution will be  enhanced due to the large values of $\mu$
associated with such low values of $\tan \beta$ and, consequently,
 the mixing may be
sufficiently large to yield a tachyonic solution \cite{COPW}-\cite{NA2}.
Thus, depending on
the case considered for the
soft supersymmetry breaking scalar masses and its
hierarchy with respect to $M_{1/2}$, as well as on the sign of $\mu$,
important constraints on the parameter space may be obtained.

For example, if we consider first the universal case  with a common
scalar mass $m_0$, for $\tan \beta = 1.2$,
which implies $M_t \simeq  160$ GeV
and  for which the value of the
 supersymmetric mass parameter  $\mu^2 \simeq 4 m_0^2 + 12 M_{1/2}^2$,
it is straighforward to show that, if one considers
the regime $M_{1/2}^2 \ll m_0^2$ for both signs of $\mu$
(or $M_{1/2}^2 > m_0^2$ for $\mu > 0$), then
a tachyon state will develop unless  $M_{1/2}
\geq \; m_t$.
For $M_{1/2}^2 > m_0^2$ and  $\mu < 0$, since there is  a partial
cancellation of the off-diagonal term which  suppresses the mixing,
no tachyonic solution may develop and, hence, no constraint
is derived from these considerations.
However, as we shall show  below,
restrictions coming from the Higgs sector will constrain this  region
of parameter space as well. Observe that for these low values of
$\tan\beta$, the necessary and sufficient conditions to avoid a
color breaking minima, Eqs. (\ref{eq:cond1}), (\ref{eq:cond2})
and (\ref{eq:cond3}), put strong restrictions on the solutions
with large left--right stop mixing.
For slightly larger values of $\tan \beta \simeq 1.8$, which correspond
to much larger values of the top quark mass, $M_t \simeq 180 GeV$, the
value of $\mu^2 \simeq 1.2 m_0^2 + 5.3 M_{1/2}^2$
is sufficiently
small so that, helped by the factor $1/\tan \beta$
appearing in the off--diagonal terms in Eq.(\ref{eq:stopmat}), there is
no possibility for a tachyon to develop  in this case and, hence, no
constraints on $M_{1/2}$ are obtained. Of course, this result holds for
larger values of    $\tan \beta$ as well.
It is interesting to notice that, although there is no necessity to be
concerned about tachyons for values of $\tan \beta \simeq 1.8$,
it is still possible
 to have light stops, $m_{\tilde{t}_1} < 150$ GeV,
 if the value of $M_{1/2} \leq 100$ GeV.
For larger values of $\tan \beta$ ($M_t > 185$ GeV)
a light stop is not possible
any longer in the universal case.

Figure 1 shows the dependence of the stop mass on the chargino mass
in the case of universal scalar masses at $M_{GUT}$,
for four different values of the top quark
mass. For low values of $M_t \leq 165$ GeV, the color breaking
constraints forbids large mixing in the stop sector and, due to
the behaviour of $m_U^2 \simeq 4 M_{1/2}^2$, a strong correlation
between the lightest stop and the lightest chargino is observed. For
larger values of $M_t$, larger mixing is allowed and a clear
distinction between the two signs of $\mu$ is observed. This distinction
is particularly clear for $M_t \simeq 175$ GeV ($\tan\beta \simeq 1.5$)
and disappears for larger values of $\tan\beta$. Observe that just for
the interesting region 165 GeV $\leq M_t \leq$ 185 GeV both light
stops and light charginos ($m_{\tilde{\chi}_1} < 70$ GeV) are allowed.
For larger (lower) values of $M_t$, it becomes more difficult to
get light stops (charginos).
Light stops and charginos are  very interesting, both for
direct experimental searches \cite{LNZ}
as well as for indirect searches
through deviations
from the Standard Model predictions for  the leptonic and hadronic
variables measured at LEP (see below).

If we consider  the non--universal case with
$m_{H_1}^2(0)$ = 0 and
$m_{H_2}^2(0)/2 = m_Q^2(0) = m_0^2/2 $
(case I), then if $M_{1/2}$
dominates the supersymmetry breaking, the
constraints coming from the requirement of avoiding a
very small stop mass are equivalent to the ones obtained in the
case of universal mass parameters. If $M_{1/2}$ is much smaller than
the parameter $m_0$, instead, an upper bound on the scalar mass
parameter is obtained,
$m_0^2 < 6 m_t^2$.
More generally, in the regime of large values of $m_0$ and
moderate values of $M_{1/2}$, it follows that
for positive values of $\mu$, in order to avoid
tachyons,
\begin{equation}
m_t^2 > 0.5 \left[ K M_{1/2}^2 +
\sqrt{ \left(K M_{1/2}^2 \right)^2 - 96 M_{1/2}^4
+ \left( \frac{m_0^2}{3} \right)^2 + \frac{4}{3} m_0^2
M_{1/2}^2 } \right]
\label{eq:fillin}
\end{equation}
with $K \simeq 10, 4.5, 0.8$ for $M_t \simeq 165, 175, 185$ GeV
($\tan\beta \simeq 1.3, 1.5, 1.9$).
For $M_t \leq 160$ GeV ($\tan\beta \leq 1.2$) this condition cannot be
fulfilled since for values of $M_{1/2}$ consistent with the present
experimental bounds on the chargino and gluino masses,
already the $M_{1/2}$ dependent part violates the
above bound. For $M_t = 165 $ GeV there is a small region for which
$M_{1/2}$ is rather small, $m_0$ is rather large and for which this
condition is fulfilled (see Fig. 2).
For negative values of $\mu$, the off--diagonal terms
are small and with a very weak dependence on the soft breaking parameters
$m_0$ and $M_{1/2}$. Hence, one obtains a bound which is basically
equivalent to the positivity of the diagonal term,
\begin{equation}
m_U^2 > - (m_t^2 + D_{t_R}).
\label{eq:posdiag}
\end{equation}
In the present case, Eq.(\ref{eq:posdiag})  is equivalent to
\begin{equation}
m_0^2 < 6 \left( m_t^2 + D_{t_R} + 4 M_{1/2}^2 \right).
\label{eq:m0bound}
\end{equation}
Quite generally, the condition of absence of
color breaking, derived from Eq. (\ref{eq:minimu}) assures the
fulfillment of Eq.(\ref{eq:posdiag}) since the
actual bound coming from the absence of color breaking
is stronger than the one implied by Eq. (\ref{eq:m0bound}).

Fig. 2 shows the dependence of the lightest stop mass
on the lightest chargino mass
for case I of non--universal mass parameters at $M_{GUT}$
and four different values of the top quark mass.
For low values of $M_t \leq 160$ GeV, light charginos are not
allowed. This is due to the Higgs bounds and the impossibility
of getting large radiative corrections due to the bounds on
$m_0$ derived from constraints in the stop sector and the
absence of a color breaking minimum. For $M_t \simeq 165$ GeV,
there is a regime with light charginos and $\mu > 0$,
for which Eq. (\ref{eq:fillin}) and the Higgs mass bounds are fulfilled.
Apart from this region, light charginos do not appear in the spectrum
for this low values of $\tan\beta$. For $M_t \geq 175$ GeV, there
is a clear distinction between positive values of $\mu$ (lower
$m_{\tilde{t}_1}$) and negative values of $\mu$ (larger $m_{\tilde{t}_1}$).
As can be seen from figure 5, for negative values of $\mu$ and
$M_t < 185$ GeV, light charginos are not allowed,
due to the constraints in the Higgs sector.

Finally, for the last condition under study, for which
$m_{H_2}^2(0) = 0$ and $m_{H_1}^2(0) = 2 m_Q^2(0) \equiv m_0^2$
(case II),
the requirement of absence of tachyons in the stop sector differs
from the other two cases only in the limit of large values of
the soft supersymmetry breaking terms for the scalar fields at
the grand unification scale. Since now both $m_Q^2$ and $m_U^2$
grow for large values of $m_0^2$, low values of the top
squark masses may only be achieved for large values of the
left--right mixing, which naturally arise in the low $\tan\beta$
regime. In particular, for $\tan\beta \simeq 1.3$, which
approximately corresponds to $M_t \simeq 165$ GeV, and low values
of the common gaugino mass $M_{1/2}$, in order to avoid problems
in the stop spectrum,
it is necessary to have $m_0^2 \geq 0.2 m_t^2$. This bound
becomes stronger for lower values of $\tan\beta$. On the contrary,
for large values of the top quark mass,
$M_t \geq 175$ GeV, no bound on $m_0$ is obtained from these
considerations.
Figure 3 shows the dependence of the
lightest stop quark mass on the lightest chargino
mass for case II and four different top quark mass values. For low
values of $M_t \leq 165$ GeV, the colour breaking constraints are
sufficiently
strong  to put restrictions on large values of $m_0$,
particularly for low values of the chargino mass ($M_{1/2}$)
and positive values of $\mu$. For larger values of $M_t$,
there is again a distinction between positive and negative values of
$\mu$. Observe that, due to the larger mixing, lower values of the
lightest stop are always more easily obtained for positive values of
$\mu$.

\subsection{Higgs Spectrum}

Other  important features  of the spectrum at the infrared fixed point
are associated with the Higgs sector.
The Higgs spectrum is composed by three neutral scalar states --two
CP-even, h and H,  and  one CP-odd, A,  and two charged scalar states
H$^{\pm}$. Considering the one loop leading order corrections to the
running of the quartic couplings --those proportional to $m_t^4$-- and
neglecting in a first approximation the squark mixing, the masses of the
scalar states are given by,
\bea
m^2_{h,H} &=& \frac{1}{2}
\left[m_A^2 + M_Z^2 + \omega_t  \right. \nonumber \\
&  & \left. \pm \sqrt{\left(m_A^2 + M_Z^2 \right)^2
+ \omega_t^2 - 4 m_A^2 M_Z^2 \cos^2(2 \beta) + 2 \omega_t
\cos(2 \beta) \left(m_A^2 - M_Z^2 \right)} \right]
\label{eq:mhH}   \\
\nonumber \\
m_A^2  & = & m_1^2 + m_2^2  + \frac{\omega_t}{2}
\nonumber \\
& =&
 \left[ m_{H_1}^2(0) + \left(
\frac{ 2 m_Q^2(0) -  m_{H_2}^2(0)}{2} \right)
 + 4 M_{1/2}^2 -
\frac{\omega_t}{2} \right] \frac{(1+ \tan^2 \beta)}{(\tan^2 \beta -1)}
\nonumber\\
\label{eq:mA}      \\
m_{H^{\pm}}^2 & =  &  m_A^2 + M_W^2    \; .
\label{eq:mHpm}
\eea
In the above,
we  have omitted the one loop contributions proportional to
 $\omega_t/ m_t^2$, since for $\tan\beta > 1$
they are negligible with respect to the other contributions.
{}From Eq. (\ref{eq:mA}) it follows that, for lower values of $\tan \beta$,
the value  of the CP-odd eigenstate mass is enhanced.
Moreover,
larger values of $m_A$, implies as well that the charged Higgs and
the heaviest CP-even Higgs  becomes  heavier in such regime.
Indeed,
for low values of $\tan\beta \leq 2$ ($M_t \leq 190$ GeV) and
for the experimentally allowed range for the
other mass parameters, the CP-odd Higgs is always heavier than
150 GeV. In this regime, the radiative corrections give only
a relevant contribution to the lightest CP--even Higgs mass,
Eq. (\ref{eq:mhH}). In fact, for these large values of $m_A$,
$m_h$ acquires values close to its upper bound, which is independent
of the exact value of the
CP--odd mass \cite{Nir}--\cite{LNan}:

\be
(m_h^{max})^2 = M_Z^2 \cos^2(2\beta) + \frac{3}{4 \pi^2}
\frac {m_t^4}{v^2} \left[ \ln \left(
\frac{m_{\tilde{t}_1}   m_{\tilde{t}_2}}{m_t^2} \right) +
\Delta_{\theta_{\tilde{t}}} \right]
\label{eq:mhmax}
\ee
In the above, we have now considered the expression in the
case of non-negligible squark mixing
$\Delta_{\theta_{\tilde{t}}}$ is a function which
depends on the left-right mixing angle  of
the stop sector and it vanishes in the limit in which the two mass
eigenstates are equal:
$m_{\tilde{t}_1} =  m_{\tilde{t}_2}$ \cite{Nir}-\cite{LNan},
\begin{eqnarray}
\Delta_{\theta_{\tilde{t}}} & = &\left( m_{\tilde{t}_1}^2 -
m_{\tilde{t}_2}^2 \right) \frac{\sin^2 2\theta_{\tilde{t}}}
{2 m_t^2} \log \left( \frac{m_{\tilde{t}_1}^2}{m_{\tilde{t}_2}^2}
\right)
\nonumber\\
& + & \left( m_{\tilde{t}_1}^2 - m_{\tilde{t}_2}^2 \right)^2
\left( \frac{\sin^2 2 \theta_{\tilde{t}}}{4 m_t^2} \right)^2
\left[ 2 -
\frac{m_{\tilde{t}_1}^2 + m_{\tilde{t}_2}^2}
{m_{\tilde{t}_1}^2 - m_{\tilde{t}_2}^2} \log \left(
\frac{m_{\tilde{t}_1}^2}{m_{\tilde{t}_2}^2}  \right) \right],
\end{eqnarray}
where $\theta_{\tilde{t}}$ is the stop mixing angle.

Furthermore,
the infrared fixed point solution for the top quark mass has
explicit  important
implications for the lightest Higgs mass. For a given value of the
physical top quark mass, the infrared fixed point solution  is associated
with the minimun value of $\tan \beta$ compatible  with the perturbative
consistency of the theory. For values of $\tan \beta\geq 1$, lower values
of $\tan \beta$ correspond to lower values of the tree level lightest
CP-even mass, $m_h^{tree} = M_Z |cos 2 \beta|$. Therefore, the infrared
fixed point solution minimizes the tree level contribution and after
the inclusion of the radiative corrections it still gives the lowest
possible value of $m_h$ for a fixed value of $M_t$ \cite{COPW},
\cite{BABE},\cite{Cartalk}, \cite{CEQR}.
This property is very
appealing, in particular, in relation to  future Higgs searches at LEP2,
as we shall show explicitly below. In figure 4 we present the values of
the lightest Higgs mass as a function of the top quark mass at its
infrared fixed point solution, for the case of universal boundary
conditions,
and performing a scanning over the
mass parameters up to low energy squark masses of the order of 1 TeV.
For comparison, we present the upper bounds on the Higgs mass which
is obtained for larger values of $\tan\beta$. Observe that, for
$M_t \leq 175$ GeV, there is approximately a difference of 30 GeV
between the upper bound at and away from the top quark mass fixed
point. As we shall discuss below in more detail,
although the characteristic of the Higgs spectrum depend
on the boundary conditions at the grand unification
scale, these upper bounds  have a more  general validity.

In general, the lightest CP-even Higgs mass spectrum is a reflection
of the characteristics of the stop spectrum presented in Figs. 1 - 3.
For the same chargino mass, larger Higgs mass values are obtained for
positive values of $\mu$ than for negative values of $\mu$.
Quite generally,
the upper bound on the Higgs mass does not depend on the
different structure of the boundary
conditions of the scalar mass parameters at the grand unification
scale. It reads $m_h \leq$ 90 (105) (120) GeV, for $M_t \leq 165
\; (175) \; (185)$ GeV.
Fig. 5, 6 and 7 present the dependence of the lightest CP-even
Higgs on the chargino mass for four different values of the
top quark mass and for the case of universal scalar mass $m_0$
and for the cases of non--universal mass parameters
I and II, respectively.

Both the universal case and case II present similar features
and are almost indistinguishable from the point of view of the
lightest CP--even Higgs spectrum (Figures 5 and 7).
For $M_t \simeq 165$ GeV,
the Higgs mass becomes larger for a chargino mass
$m_{\tilde{\chi}_1} \simeq 100$ GeV, than for moderate values of
the chargino mass.  The Higgs mass becomes, however,
tightly bounded from above and it is always in the regime
to be tested at LEP2. For larger values of the top quark mass,
$M_t \geq 175$ GeV, and for negative (positive) values of $\mu$,
the Higgs mass lies mostly within (beyond)
the experimentally reachable regime. Observe that, even if
a light chargino is observed at LEP2, $m_{\tilde{\chi}_1^+} < 90$
GeV, nothing guarantees the observation of the lightest CP--even
Higgs, particularly for larger values of the top quark mass,
$M_t \geq 175$ GeV.

Case I (Fig. 6) is
easily distinguishable from the above two cases, due to
the more definite values of the Higgs mass related to the smaller
allowed dependence on the scalar mass parameter $m_0^2$. Observe
that, although the absolute upper bound on $m_h$ for a given $M_t$
does not significantly change, due to this particular structure
of the high energy soft supersymmetry breaking mass parameters,
the Higgs mass is in general pushed to lower values than in
the case of universal parameters.
Therefore, unlike these two  cases, the observation of
a light chargino at LEP2 would almost guarantee the observation of
a light neutral Higgs if $M_t \leq$ 185 GeV in this case.

\section{Precision Data Variables}

In this section we shall define the experimental variables,
which we shall use to
analyze the implications of the
infrared fixed point solution for the precision data
analysis at LEP.  In particular, we will follow the
procedure  of Altarelli, Barbieri, Caravaglios
and Jadach \cite{ABJ}-\cite{ABC2},
which consist in
parametrizing  the electroweak radiative corrections
in terms of four parameters: $\epsilon_1$, which is
directly related to the $Z$-boson lepton width and is
closely related to the parameter $\Delta\rho(0)$
usually defined in the literature \cite{Veltman}(see below),
the parameter $\epsilon_2$, which is related to the parameter
$\Delta r_W$, which measures the radiative corrections to
the $W^{\pm}$ boson masses \cite{WS},
the parameter $\epsilon_3$,
closely related to the radiative corrections to the weak
mixing angle, and the parameter $\epsilon_b$, which is
related to the radiative corrections to the $Z$-$b \bar{b}$
vertex \cite{epsbst}-\cite{Gautam}.
In this work, we shall concentrate on the parameters
$\epsilon_1$ and $\epsilon_b$ which are the only ones which
keep a quadratic dependence on the top quark mass.
This parametrization is based on
the precise knowledge of $G_F$, $\alpha$ and $M_Z^2$, which are
used as a basis for the precision data analysis.

The variable $\epsilon_1$
may be directly obtained from the measurements of the
$Z$--boson width and the forward--backward lepton asymmetries.
Indeed the forward backward asymmetries may be parametrized in
terms of the renormalized vector and axial lepton--$Z$ boson
couplings, $g_V$ and $g_A$ in the following way:
\begin{equation}
A_{FB}^l = \frac{ 3 (g_V/g_A)^2 }{\left[ 1 +
(g_V/g_A)^2 \right]^2 }.
\end{equation}
{}From $g_V/g_A$ is it possible to define an effective
weak mixing angle
\begin{eqnarray}
\frac{g_V}{g_A} & = & 1 - 4 \sin^2 \theta_W^{eff}
\nonumber\\
& = & 1 - 4 \left( 1 + \Delta k \right) s_0^2,
\label{eq:gvga}
\end{eqnarray}
where
\begin{equation}
s_0^2 c_0^2 = \frac{ \pi \alpha(M_Z) }{ \sqrt{2} G_F M_Z^2 }.
\end{equation}
$\Delta k$ is a measure of the radiative corrections
to the weak mixing angle, which are quadratically dependent on
the top quark mass. Observe that
the angle $s_0^2$ already contains the
running between low energies and the energy scale $M_Z$.

The total leptonic width may be also parametrized in terms of the
axial and vector lepton couplings,
\begin{equation}
\Gamma_l = \frac{ G M_Z^3}{ 6 \pi \sqrt{2} }  g_A^2
\left( 1 + \frac{ g_V^2 }{g_A^2}
\right)
\end{equation}
{}From the knowledge of the asymmetris and the lepton width one can
obtain the axial coupling
\begin{equation}
g_A^2  = \frac{1}{4} \left( 1 + \Delta\rho \right).
\label{eq:ga2}
\end{equation}

Then, the variable $\epsilon_1 \equiv \Delta \rho$
receives four different contributions\cite{BFC}:
\begin{equation}
\epsilon_1 = e_1 - e_5 - \frac{\delta G}{G} - 4 \delta{g_A},
\end{equation}
where  $e_1 \equiv \Delta \rho(0)$ is given by,
\begin{equation}
e_1 = \frac{ \Pi_{33}(0) - \Pi_{WW}(0) }{M_W^2},
\end{equation}
with $\Pi_{33}(0)$ and $\Pi_{WW}(0)$  the
zero momentum vacuum polarization
contributions to the $W_3$ and $W^{\pm}$ gauge bosons. In general,
\begin{equation}
\Pi_{i j}^{\mu\nu}(q) = -i g^{\mu\nu} \Pi_(q^2) + q^{\mu} q^{\nu}
terms.
\end{equation}
with $i,j = W,\gamma,Z$ or $i,j$ = 0,3 for the $W_3$ or $B$ bosons,
respectively.  The term
$e_5$ proceeds from the wave function renormalization
constant of the $Z$ boson at $q^2 = M_Z^2$ and is given by
\begin{equation}
e_5 = \left\{ q^2 \left[\frac{d}{dq^2}
\frac{\left( \Pi_{ZZ}(q^2) -
\Pi_{ZZ}(0) \right)}{q^2} \right] \right\}_{q^2 = M_Z^2}.
\label{eq:e5}
\end{equation}
The contribution of
$e_1$ and $e_5$ include all the dominant vacuum polarization
effects to the renormalized coupling $g_A$.
Finally, the vertex and box corrections are included in the
variables $\delta g_A$ and $\delta G/G$, as described, for example,
in Ref. \cite{BFC}.
The dominant contributions to the $\epsilon_1$ parameter are
described in Appendix A.

Using Eqs. (\ref{eq:gvga})--(\ref{eq:ga2}) and the precise values
for $G_F$, $\alpha(M_Z)$ and $M_Z$ in the standard model,
the variable $\epsilon_1$ is related with the
asymmetries and the $Z$--boson leptonic width through the
following expression \cite{Altatalk},
\begin{equation}
\epsilon_1 =  - 0.9882 + 0.01196 \frac{\Gamma_l}{MeV} -
0.1511 \frac{g_V}{g_A}.
\end{equation}

Analogously to the variable $\epsilon_1$, the variable $\epsilon_b$
may be defined as a function of the axial and vector couplings of
the b--quark to the $Z^0$--boson. In the low $\tan\beta$ regime,
the relevant contributions, quadratically dependent on the top
quark mass,  may be analysed in terms of only the coupling of the
left handed bottom quarks to the $Z^0$ gauge boson.
Formally, in this regime
$\epsilon_b$ is defined from the relation
\begin{equation}
g_A^b = -\frac{1}{2} \left( 1 + \frac{\Delta \rho}{2} \right)
\left( 1 + \epsilon_b \right),
\end{equation}
with
\begin{equation}
g_L^b = \left(-\frac{1}{2} + \frac{1}{3} \sin^2\theta^{eff}_W
- \frac{\epsilon_b}{2} \right) \left( 1 + \frac{\Delta\rho}{2}
\right),
\end{equation}
and
\begin{equation}
g_R^b = \frac{\sin^2\theta^{eff}_W}{3} \left( 1 + \frac{\Delta\rho}{2}
\right).
\end{equation}
Experimentally,
the variable $\epsilon_b$ can be best obtained from the
ratio of the $Z \rightarrow b \bar{b}$ width to the total hadronic
width. It can be shown that the branching ratio is given by \cite{Altatalk}
\begin{equation}
\frac{\Gamma_b}{\Gamma_h} \simeq 0.2182 \left[ 1 + 1.79 \epsilon_b
- 0.06 \epsilon_1 + 0.07 \epsilon_3 \right],
\end{equation}
where the variable $\epsilon_3$ is defined as
\begin{equation}
\epsilon_3 = c_0^2 \Delta\rho + \left(c_0^2 - s_0^2\right) \Delta k.
\end{equation}
and depends only logarithmically on the top quark mass.
The most relevant contributions to the variable $\epsilon_b$
in the low $\tan\beta$ regime and within the minimal supersymmetric
standard model  are described in Appendix B.

In the above, we have given the dependence of the precision data
variables, on the observables which are most sensitive
to it. From the point of view of the experimental analysis, however,
it is possible to extend the fit of the variables $\epsilon_1$,
$\epsilon_3$ and $\epsilon_b$ by the introduction of other
measured
observables. This may be performed by, for example,
including all purely leptonic quantities
at the $Z^0$--pole, or the data on the $b$--quark from the
forward--backward asymmetry, or simply to include all
observables measured at the $Z^0$ peak at the LEP experiment.
This last step may be performed by assuming that all relevant
deviations from the standard model may be associated with
either vacuum
polarization effects or  corrections to the $Z \rightarrow
b \bar{b}$ vertex. The global fit to the data reduces the dependence
on any single experiment and hence provide a more realistic estimate
of the precision data variables.
For the comparison of the theoretical
results  to the experimental data, we shall  use the values
of the variables which are obtained from these extended fits
at the 90 $\%$ confidence level.

\section{Indirect Signals of Supersymmetric Particles}

In this section we shall investigate  the
possible experimental signature of
supersymmetric particles in the variables $\epsilon_1$,
$\epsilon_b$ and the rate of the rare $b$ decay, $b \rightarrow
s \gamma$. We shall study  this in the case of universality
of the soft supersymmetry breaking parameters at the unification
scale  and in the two
characteristic cases of non-universal soft supersymmetry breaking
scalar mass
parameters discussed in sections 3 and 4 (cases I and II).
A related analysis, within the framework of
superstring--inspired $SU(5) \times U(1)$
supergravity models, was recently performed in Ref. \cite{LNPZ}.

Before analysing each case
in detail, let us summarize the most relevant supersymmetric
effects in these three experimental variables.
The main supersymmetric contributions to the variable $\epsilon_1$
comes from the chargino and stop sectors and are summarized
in Appendix A. The stop contribution is always positive, and
becomes relevant whenever there are light stops, with masses
$m_{\tilde{t}} < 300$ GeV and with a non-negligible
component in the $\tilde{t}_L$ squark.  Due to the renormalization
group behaviour of the mass parameters $m_Q^2$ and $m_U^2$,
$m_Q^2$ is always larger than $m_U^2$ at low energies (see Table 1 and
Eq. (\ref{eq:massp}) )
and, hence,
in the cases under analysis the light stop has a dominant
right handed component. A left handed component appears
mainly through the mixing, which does not strongly affect
the behaviour of $\Delta\rho(0)$ \cite{BM}.
Hence, even in the case
of light stops, with masses lower than 100 GeV, the
potentially large positive contributions to the $\rho$
parameters are in general suppressed. Light charginos,
instead, give a negative contribution to $\epsilon_1$,
which become large if the lightest chargino mass
$m_{\tilde{\chi}^+_l} < 70 $ GeV. Since in most cases, light
stops may only appear when charginos with masses close to
the present experimental bounds are present in the spectrum,
the light stop effect is in general screened by the chargino
contribution.

The main contributions to the
variable  $\epsilon_b$ in the minimal supersymmetric standard
model in the low $\tan\beta$ regime come from the standard
$W^{\pm}$--top loop,
the charged Higgs--top and the chargino--stop
loops and are summarized in Appendix B.
The charged Higgs contribution pushes $\epsilon_b$ in
the same direction as the standard model one, while the chargino
contributions tend to suppress the standard model corrections to
the $Z^0 - b \bar{b}$ vertex.
In the models under consideration, for $M_t < 185$ GeV
($\tan\beta < 2$), the charged Higgs is sufficiently heavy
to give only a moderate contribution to the $\epsilon_b$
variable. The chargino contribution, instead, may become sizeable,
particularly when light charginos and light stops are present in
the spectrum. The largest
chargino contributions, quadratically dependent on the top quark
mass, appear in the case in which the lightest stop has a
relevant right handed component and the lightest chargino has
a relevant component in the charged Higgsino. Although the first
condition is mostly
satisfied for the cases of soft supersymmetry breaking terms
under consideration, due
to the large values of $\mu$, the lightest chargino has
a dominant wino component, which reduces the supersymmetric effects
on $\epsilon_b$. Still, as we shall show, relatively large effects are
still possible.

The decay rate  $b \rightarrow s \gamma$ receives also contributions
from the standard $W^{\pm}$--top loop, the charged Higgs--top
loop and the chargino--stop loops.  The predictions for
this decay rate within the Standard Model has been recently
analysed by several authors \cite{bsganal}. A general expression for
the supersymmetric contributions has been presented in
Refs. \cite{BBMR} and \cite{BG2}, and we shall not rewrite it here.
The relevant properties are the following:
As in the case of $\epsilon_b$, the charged
Higgs contribution tends to enhance the standard model signal.
In the supersymmetric limit,
$\mu = 0$ and $\tan\beta = 1$, the stop-chargino
contribution exactly cancels
the charged and standard model ones and the total rate is zero.
Although for the
experimentally preferred values of $M_t \simeq 174 \pm 17$ GeV \cite{CDF}
cases under
consideration, its infrared fixed point solution yields values of
$\tan\beta$ close to one, the values of the sparticle masses are
far away from their supersymmetric expressions.
Indeed, large values of $\mu$ are obtained
and the soft supersymmetry breaking terms are in general not
negligible. Furthermore,
in the cases analysed in this work,
the supersymmetric contribution singles out the
sign of the mass parameter $\mu$. For
moderate positive values of
$\mu$ there is a large suppression of the standard model
decay rate, while for moderate not negative values of $\mu$
the branching ratio tends to be enhanced
(The dependence on the sign of $\mu$  is stronger  in the large $\tan\beta$
regime, $\tan\beta \geq 30$
\cite{bsganal},\cite{wefour}, which will not be analysed
in the present work).  In addition, as we discussed section 5,
for positive values of
$\mu$, due to the larger values of the stop mixing, it is easier to
obtain smaller stop masses without being in conflict with the
experimental bounds on the lightest
CP-even Higgs mass  ($m_h \geq 60$ GeV, for $m_A \geq 150$
GeV). For a given value of $M_t$ ($\tan\beta$) larger values
of $\mu$ are associated with heavy sparticles and hence the
Standard Model decay rate tends to be recovered (see figure 12).

\subsection{Dependence of the precision data variables on the light
chargino mass}

Figure 8 shows the dependence of the parameter $\epsilon_1$ for
the case of universal scalar masses at the grand unification scale
and for three different values of the top quark mass.
We observe that the qualitative features do not depend on the
top quark mass: A departure from the Standard Model prediction occurs
only for light chargino masses $m_{\tilde{\chi}_l^+} < 100$ GeV.
As we discussed above, due to the small left handed component of the
lightest stop, the main contribution is mainly negative, and this
remain a general feature independently of the exact value of the top
quark mass. Comparing the
theoretical predictions with the recent fit to the LEP
and SLD data \cite{Alfit},
\begin{equation}
\epsilon_1 = ( 3.5 \pm 2.9 ) \times 10^{-3}
\label{eq:eps1}
\end{equation}
at the 90 $\%$ confidence level (1.64 standard deviations), we see
that, while light charginos with masses $m_{\tilde{\chi}_1^+} < 70$ GeV
are not in conflict with the present data, only for large values of
the top quark mass $M_t \geq 185$ GeV, are they preferred to heavier
ones. On the other hand, very light charginos, with masses
$m_{\tilde{\chi}_1} \leq 50$ GeV are disfavoured by the present data.
It is important to remind the reader, however, that the present
analysis looses its validity for $m_{\tilde{\chi}_1} \leq 48$ GeV and
hence the predictions for chargino masses lower than 50 GeV cannot
be fully trusted.

For the case of non--universal soft supersymmetry  breaking
scalar mass parameters
at the grand unification scale  the main features of
the universal case are preserved. In figures 9 and 10 we show the
dependence of $\epsilon_1$ on the chargino mass for the cases I and II,
respectively, and a top quark mass $M_t = 175$ GeV. We see that, in
spite of the quite different characteristics of the stop spectrum with
respect to the universal case
(see figures 1--3), no significant difference
is observed with respect to the behaviour depicted in figure 8.

In figure 11 we present the behaviour of $\epsilon_b$ as
a function of the lightest chargino mass for the case of universality
of the soft supersymmetry breaking parameters  and for three
different values of the top quark mass. As in the case of $\epsilon_1$,
a significant departure from the Standard Model predictions may only
be observed if the lightest chargino mass acquires rather small values,
$m_{\tilde{\chi}_1} < 100$ GeV. However, in the presence of light
charginos, the supersymmetric predictions for $\epsilon_b$ show
a larger spreading of values for the
precision data variable $\epsilon_b$ than for $\epsilon_1$.
This is related to the dependence of
$\epsilon_b$ on the lightest stop mass. Indeed, due to the large
component of the lightest stop on the right handed top squark,
$\epsilon_b$  gets significantly changed for lower values of
$m_{\tilde{t}_1}$.  Since for $M_t \geq 175$ GeV, light stops are
only possible for $\mu > 0$, the largest values of $\epsilon_b$
are associated with positive values of $\mu$. Taking into account the
recent fit to the variable $\epsilon_b$ \cite{Alfit},
\begin{equation}
\epsilon_b = ( 0.9 \pm 6.8 ) \times 10^{-3}
\label{eq:epsb}
\end{equation}
at the $90 \%$ confidence level, we see that the present data tends
to prefer a light chargino, with mass $m_{\tilde{\chi}_1} \leq 80$ GeV.
Observe that, for $\epsilon_b$ we take the fit to all LEP
and SLD data,
instead of taking the particular value obtained from the partial
width $\Gamma_b/\Gamma_h$. If we just fit $\epsilon_b$ with this last
variable according to the last reported data, $\Gamma_b / \Gamma_h =
0.2202 \pm 0.0020$ \cite{Schaile}
we would get a larger central value, but also a
larger error at the $90 \%$ confidence level, $\epsilon_b =
(5.1 \pm 8.4)\; 10^{-3}$.  Tighter bounds on the spectrum would
be obtained if we took this latter value to perform our analysis.

Figure 10 shows the dependence of $\epsilon_b$ as a function of the
lightest chargino mass for case II and a top quark mass $M_t = 175$
GeV. The characteristic features of this case are similar to the
case of universal soft supersymmetry breaking parameters.
Only a smaller concentration of points with larger values of
$\epsilon_b$ is observed, related to the larger values of
$m_U^2$ for the same value of the chargino mass parameters (see
Table 1), which imply a smaller
right handed component of the lightest stop.

Finally, in case I, larger values of the variable
$\epsilon_b$ than in the other two cases may be obtained.
In Figure 9 we display the corresponding
dependence of $\epsilon_b$ as a function
of the chargino mass for this case, with a top quark mass $M_t =
175$ GeV. Observe that values of $\epsilon_b$ close to zero are
ppossible in this case.
This is due to the fact that the right handed stop mass parameter
$m_U^2$ can take very small values and the right handed component
of the lightest stop increases. In principle, a light stop
may be obtained in this case for sufficiently low values of
$m_U^2$, even when the mixing is negligible.  However, large negative
values of $m_U^2$ induce an unacceptable color breaking minimum,
Eq. (\ref{eq:minimu}), and hence light stops and larger values of
$\epsilon_b$ are only possible for positive values of $\mu$, as
in the universal case. Again, agreement of the theoretical results with
the present experimental data at the 90 $\%$ confidence level
may only be obtained for sufficiently light charginos
$m_{\tilde{\chi}_1^+} < 70$ GeV.

 \section{On the $b \rightarrow s \gamma$ decay rate}

In figure 12 we present the behaviour of the ratio of the
prediction for the
decay rate $b \rightarrow s \gamma$
to the standard model one, as a function of the
supersymmetric mass parameter $\mu$ for the universal case
and  for three different values of the top quark mass $M_t$.
As we discussed before, a clear dependence of $b \rightarrow
s \gamma$ on the sign of $\mu$ is observed.
For negative values of $\mu$ and a fixed value of the
top quark mass $M_t \leq 175$
GeV, most of the theoretical predictions are close to the
Standard Model ones, with a decay rate varying between 0.7 and
1.4 times the Standard Model prediction. The maximum departure is
always noticed for the smallest values of $|\mu|$, associated with a
light spectrum. For larger values of the top quark mas $M_t \geq 185$
GeV, a larger departure is possible, with a relative decay rate which
may be close to two. Recently, the experimental value
of the $b \rightarrow s \gamma$ decay branching ratio has
been reported \cite{bsgexp},
\begin{equation}
BR(b \rightarrow s \gamma) = (2.32 \pm 0.97) \times
(1 \pm 0.15) \times \left[1 - (M_b - 4.87)\right] \times
\; 10^{-4}
\label{eq:bsg}
\end{equation}
where the second error is  systematical,
the bottom mass is given in GeV, and all errors have been treated
at the $90 \%$ confidence level ($1.64 \sigma$ deviations).
The above range
allows, in principle,
to put constraints in the supersymmetric spectrum.
There are, however,
large theoretical uncertainties associated with the standard
model predictions, which for a top quark mass in the range
$M_t \simeq 165$--185 GeV, and
at the $90 \%$ confidence level reads \cite{Buras},
\begin{equation}
BR(b \rightarrow s \gamma) (SM) \simeq
( 3.1 \pm 1.5 ) \times 10^{-4},
\end{equation}
with a small dependence of the central value on the top quark
mass ($\Delta BR(b \rightarrow s \gamma \simeq \pm 0.1 \; 10^{-4}$),
which is negligible in comparison to the theoretical error
associated with QCD uncertainties.
Hence, the presently allowed values for the relative decay rate
at the 90 $\%$ confidence level translates into:
\begin{equation}
0.25 \leq
\frac{BR(b \rightarrow s \gamma)}{BR(b \rightarrow s \gamma) (SM)}
\leq 2.5,
\end{equation}
Observe that, to obtain the allowed
range, we have minimized the theoretical uncertainty related to the
bottom mass ($M_b = 4.9 \pm 0.3$ GeV) \cite{Partd}.
Had we included this uncertainty, the range would be
slightly larger than the one considered above. We believe, however,
that the above gives a conservative estimate of the experimental
values allowed at present, and it agrees  quantitatively
level  with the one reported in Ref. \cite{bsgexp}.
Hence, the relatively large values of the decay rate obtained for
$M_t \simeq 185$ GeV  are still  acceptable when all uncertainties
are taken into account.

For positive values of $\mu$, instead, the supersymmetric model tends
to predict values of the decay rate smaller than in the Standard
Model. The lower values of the stop mass associated with positive
values of $\mu$ ( and hence, with a larger mixing )
contribute to this behaviour, since they enhance the negative
chargino--stop loop contributions. In fact,
for $M_t \leq 175$ GeV, both  stops
and charginos may be sufficiently light and the
$b \rightarrow s \gamma$ decay rate may acquire very low values.
As in the case of negative values of $\mu$, however,
apart from a few solutions for $M_t \simeq 165$ GeV, the present
uncertainties do not allow  to put strong bounds on these models for
any of the values of $M_t$ considered in Fig. 12.

For the case of non--universal parameters at the grand unification
scale, cases I and II, the qualitative behaviour is the same
as in the case of universal soft supersymmetry breaking parameters.
In figures 9 and 10 we present the results for the
relative decay rate as a function of $\mu$ for cases I and II,
respectively, and a top quark mass $M_t = 175$ GeV. In case
I lower values of the relative decay rate than in the universal
case are possible for positive values of $\mu$, and some of the
predictions lie outside the experimentally allowed range. Due to
the weak dependence of $\mu$ on $m_0$, $\mu$ is strongly correlated
with the lightest chargino mass in this case, and hence, the
solutions, which are experimentally excluded
by these considerations, correspond to very
light chargino mass values.  As we shall see in the  section 9, these
are just the solutions
which tend to give larger values of $\epsilon_b$.  In case II,
instead, the theoretical predicted range is similar to the one
predicted in
the case of universal mass parameters,
and $b \rightarrow s \gamma$ remains in the
experimentally allowed range for all acceptable values of $\mu$.

\section{Correlated  fit to the Data}

In section 8, we present the theoretical predictions for
different experimental variables as a function of  relevant
supersymmetric mass parameters. However, we did not discuss the
correlations between the different variables, which become essential
at the point of considering the experimentally allowed models.
For instance, models with a value of $\epsilon_b$ closer
to the present experimental central value may be in conflict with
either the bounds on $b \rightarrow s \gamma$ or, since they are
always obtained in the presence of light charginos, they may be
in conflict with the present bounds on the $\epsilon_1$ variable.
It is the purpose of this section to analyze  these correlations.

In figure 13 we give the correlation between $\epsilon_b$ and
$\epsilon_1$ for
the case of universal mass parameters
and for three different values of the
top quark mass $M_t$. We see that larger values of $\epsilon_b$
are necessarily associated with relatively smaller
values of $\epsilon_1$,
although for $M_t \leq 175$ GeV there are a few  solutions for
which $\epsilon_1$ remains at moderate values ($\epsilon_1 \simeq
1$--$2 \times 10^{-3}$)
and $\epsilon_b$ is relatively large
($\epsilon_b  \simeq -3 \times 10^{-3}$).
These solutions are associated with light stops ($m_{\tilde{t}_1}
< 150$ GeV) and light charginos ($m_{\tilde{\chi}_1} < 70$ GeV), which
are not too close to the $Z^0$ boson mass threshold. For
$M_t \geq 185$ GeV, stops are heavy and all solutions lie beyond the
present $90 \%$ confidence level for $\epsilon_b$.  In fact,
not only the standard model prediction further decreases with respect
to lower top quark masses, but also the
deviations with respect to the standard model prediction are smaller
in this case. The variable $\epsilon_1$, instead, can vary within a
large range of values, depending on the lightest chargino mass.

In figure 14
we show the correlation between $\epsilon_b$ and $\epsilon_1$ for
cases I and II and for a top quark mass $M_t = 175$ GeV. Most of the
properties of the case with universal mass parameters
are preserved in these two cases. However, for
acceptable values of $\epsilon_1$, larger values of $\epsilon_b$
may be obtained in case I, while in case II smaller
values of $\epsilon_b$ predicteded. These properties may be
easily understood from the characteristics of the stop and
chargino spectra shown in figures 1--3. Observe that, values of
$\epsilon_b \simeq -2 \; 10^{-3}$
may be obtained in case I for acceptable
values of $\epsilon_1 \simeq 1$--$2 \; 10^{-3}$.
Observe that due to the behaviour of $\epsilon_1$ for
chargino masses $m_{\tilde{\chi}_1^+}$ very close to their
production threshold at the $Z^0$ peak (see figure 8), our
scanning shows  few solutions in figure 14  for values of
$\epsilon_1 \leq 2 \times 10^{-3}$. To fill that area with
solutions woud demand a very fine scanning for values of
$m_{\tilde{\chi}^+} < 60$ GeV.

Also interesting is the correlation between $\epsilon_b$ and
$b \rightarrow s \gamma$, which we depict in figure 15
for the case of universal mass paramters at $M_{GUT}$
and three values of the top quark mass. For negative values of
$\mu$ (see also figure 12),
larger values of $\epsilon_b$ are only possible for
$M_t \leq 165$ GeV, for which perfectly acceptable values of
$b \rightarrow s \gamma$ are obtained. Observe, however, that
for $M_t \simeq 165$ GeV,
the combination of the bounds on $\epsilon_b$, $\epsilon_1$ and
$b \rightarrow s \gamma$ restricts
$\epsilon_b < -3.2 \times 10^{-3}$ in this case.
For $M_t \simeq 175$ GeV, $b \rightarrow s \gamma$ does not impose
additional constraint but the bounds on $\epsilon_1$ are strong
enough to constraint $\epsilon_b < -3.6 \times 10^{-3}$
in this case. Much smaller
values of $\epsilon_b$ are predicted for $M_t \geq 185$ GeV.

Figure 16 shows the correlation of $b \rightarrow s \gamma$ with
$\epsilon_b$ for the cases of non--universal mass parameters
I and II and a top quark
mass $M_t = 175$ GeV. We see that, unlike the case
of universal mass parameters,
in case I the experimental range for
$b \rightarrow s \gamma$ puts additional constraints on the
spectrum. The variable $\epsilon_b$ can still take values lower
than in the standard model, but still away from zero. For
$M_t \simeq 175$ GeV, the correlated fit leads to a value of
$\epsilon_b < - 2.5 \; 10^{-3}$. As in the case of universal mass
parameters at $M_{GUT}$,
no significant variation of this bound
is obtained for lower values of the top quark mass, while for
larger values of the top quark mass $\epsilon_b$ tends to lower
values.  Finally,
from the point
of view of the range of allowed values for the experimental
variables,
case II is equivalent to the case of universal mass parameters,
once the full experimental constraints considered
in this work are taken into account.

\section{Conclusions}

In the present work, we have analysed the theoretical predictions
for the Higgs and supersymmetric spectrum and their indirect
experimental signals at the top quark mass
infrared fixed point solution
for different boundary conditions of the scalar mass parameters
at the grand unification scale. We have shown that even though
the stop mass range significantly changes for different boundary
conditions, the predicted lightest CP-even Higgs mass range
remains unchanged, leading to rather general upper bounds for
this mass,
$m_h \leq 90 \; (105) \; (120)$ GeV for
$M_t \leq 165 \; (175) \; (185)$ GeV.
The correlation between the lightest Higgs mass and the chargino
spectrum, however, depends on the chosen
high energy boundary conditions for the mass parameters.
Interesting enough, for $M_t \geq 175$ GeV, the observation of
a light chargino at LEP2, does not guarantee the observation of the
lightest CP-even Higgs mass, particularly for positive signs of
$\mu$ for which the mixing is maximized. However,
for $M_t < 185$ GeV, light stops may
appear in the spectrum in this case.
The allowed stop
spectrum in the presence of a light chargino, strongly depends on
the high energy boundary conditions. For two of the cases considered,
the case of universal scalar mass parameters at $M_{GUT}$ and
the case I, for which the dominant dependence of the supersymmetric
mass parameter $\mu$ on the scalar mass parameters vanishes,
and a top quark mass $M_t \leq 175$ GeV,
a light chargino, with mass $m_{\tilde{\chi}_1} \leq 70$ GeV is
always
associated with a light stop, with mass $m_{\tilde{t}_1} \leq
150$ GeV.  In case II, for which the right handed stop mass parameter
increases with the supersymmetry breaking
scalar mass parameter $m_0$, heavier stops may appear together with
light charginos.

The experimental variables analysed in this work are a reflection
of the characteristics of the
Higgs and supersymmetric spectrum. The variable $\epsilon_1$
receives  a significant negative correction only  for low values of the
chargino mass $m_{\tilde{\chi}_1^+} \leq 70$ GeV. The potentially large
positive correction associated with the stop spectrum is mostly
suppressed due to the relatively small
left handed component of the lightest stop. These
properties do not strongly
depend on the different boundary conditions analysed in the present
work.  The variable $\epsilon_b$ receives also a significant
correction, with respect to the Standard Model prediction only for
sufficiently light charginos, $m_{\tilde{\chi}_1^+} < 100$ GeV. The
correction is mainly positive, rendering $\epsilon_b$
closer to the experimentally
allowed range than in the Standard Model case. Due to the large
component of the lightest stop on the right handed top squark, the
variable $\epsilon_b$ depends also on the lightest stop mass. Hence,
it is mostly larger for positive values of $\mu$, for which lighter
stops are possible, particularly for $M_t \geq 175$ GeV.
Finally, the corrections to the
decay rate $b \rightarrow s \gamma$ are also maximized in the case
of light charginos and light stops. This experimental variable has
a strong dependence on the sign of $\mu$. For positive values of
$\mu$, the prediction for the
decay rate is generally larger than the standard model one,
while for negative values of $\mu$ it is generally smaller.

In the case of universal soft
supersymmetry breaking scalar mass parameters, values of
$\epsilon_b \simeq -2 \; 10^{-3}$
may be obtained for sufficiently low values of the
chargino masses. However,
these values are achieved for very low
values of the chargino masses and are in conflict with the
experimental value of the variable
$\epsilon_1$ at the 90 $\%$ confidence level. Due to the present
theoretical uncertainties in the computation of
$BR(b\rightarrow s \gamma)$,
the recent experimental measurement of this branching ratio yields
no relevant additional constraints on the allowed mass parameters
in the case of universal mass parameters at $M_{GUT}$.
In general, for $M_t \geq 165$ GeV,  the allowed values for the
variable $\epsilon_b < -3.2 \times 10^{-3}$ in this case.
Values compatible with the
present experimental bounds on $\epsilon_b$ at the 90 $\%$ confidence
level are always associated with light charginos $m_{\tilde{\chi}_1}
< 100$ GeV and values of the variable $\epsilon_1$
which are lower
than the standard model prediction, but are mostly  consistent
with the present experimental data. In fact, the theoretical predictions
for $\epsilon_1$, within the experimentally allowed range for all
variables, reads $\epsilon_1 \simeq 0.6$--$5 \; 10^{-3}$.
The decay rate $b \rightarrow s \gamma$
stays in the experimentally acceptable range, with values which
tend to be mostly lower than in the Standard Model case.

In the case $m_{H_1}^2(0) = 0$, $m_{H_2}^2(0) = 2 m_Q^2(0)$ (case I),
many of the above discussed features are preserved, although larger
values of $\epsilon_b$ are possible. Values of $\epsilon_b \simeq 0$,
which are not in conflict with the  bounds on the spectrum, lead
however to
too low values of either $\epsilon_1$ or the
branching ratio $BR(b \rightarrow s \gamma)$.
In general, for $M_t \geq 165$ GeV, $\epsilon_b < -2.5 \; 10^{-3}$
in this case. As in the universal case,
consistency with the present experimental bounds
lead to light charginos, values of the variable
$\epsilon_1 \simeq 0.6$--$5 \times 10^{-3}$
and a $b \rightarrow s \gamma$ decay
rate, which is mostly below the standard model prediction.
Finally, in the case
$m_{H_2}^2(0) = 0$, $m_{H_1}^2(0) = 2 m_Q^2(0)$ (case II), the bounds
on the $\epsilon$ parameters
are equivalent to the ones found in the case of universal conditions
at the grand unification scale.

The discrepancy between the experimentally allowed value of $\epsilon_b$
and the standard model prediction is mostly due to the lack of
agreement of the standard model  prediction for the branching ratio
$\Gamma_b/\Gamma_h$ and the
corresponding experimental value. Indeed, the standard model
prediction for $\epsilon_b$ lies beyond the experimental value at
90 $\%$ confidence level.
The determination of this partial width is, however, a delicate
experimental problem and there are some unresolved issues
related to it. Hence, it is still premature to claim  evidence
of new physics based only on the  $\epsilon_b$ variable.
If the present
tendency is mantained after these issues are solved,
the low energy supersymmetric grand unified models have the
power of closing the gap between theory and experiment. This
will demand light charginos and light stops.
If this is the case, we should see supersymmetric particles
either at LEP2 or at the next Tevatron run. Hence, within the
phenomenologically attractive
scenario of minimal supersymmetric grand unified theories,
if the present experimental bounds on $\epsilon_b$
were mantained, the above property,
together with the tight upper bounds on the Higgs mass,
promises a potentially rich
phenomenology at present and near future colliders.
{}~\\
{}~\\
{}~\\
{\bf{Acknowledgements}} The authors would like to thank G. Altarelli,
R. Barbieri, E. Lisi and J. Sola for very interesting discussions.
This work is partially supported by the Worldlab. \\
{}~\\
{}~\\
{\bf{Note added in proof}} After this work was completed, two
independent works  have appeared \cite{KP},
in which the behaviour of the variable
$\epsilon_b$ within the minimal supersymmetric model  is
analysed.

\newpage
{}~\\
{\large{\bf {Appendix A.}}} \\
{}~\\

In this appendix we describe the largest contributions to the
 parameter $\epsilon_1$ in the minimal supersymmetric standard
model. If charginos are sufficiently heavy,
$m_{\tilde{\chi}^+_l} \geq 80 GeV$, the only large
supersymmetric contributions to the
parameter $\epsilon_1$ comes from the stop--sbottom
sector.  This contribution is analogous to the dominant one
coming from the top--bottom left handed multiplet, which reads,
\begin{equation}
\epsilon_1^{t-b} = \frac{3 \; \alpha}{16 \pi \; \sin^2\theta_W
\; M_W^2} \left[ M_t^2 + M_b^2 - \frac{2 M_t^2 M_b^2}{M_t^2
- M_b^2} \ln \left( \frac{M_t^2}{M_b^2} \right) \right].
\label{eq:eps1mt}
\end{equation}
Due to the large hierarchy between the top and the bottom masses,
the above expression, Eq. (\ref{eq:eps1mt}) is completely
dominated by the first term inside the bracket.

Concerning the stop--sbottom sector,
in principle, only the  supersymmetric partners
of the left handed top and bottom quarks contribute to
$\epsilon_1$. However, due to the squark mixing governed
by the $A_t$ and $\mu$ parameters, these are not the mass
eigenstates of the model. In terms of the mass
eigenstates $m_{\tilde{t}_{1,2}}$ and $m_{\tilde{b}_{1,2}}$
the dominant stop - sbottom contribution to $\epsilon_1$
is given by \cite{WS}
\begin{eqnarray}
\epsilon_1^{\tilde{t}-\tilde{b}} & = &
\frac{3 \; \alpha}{16 \pi  \sin^2\theta_W
\; M_W^2} \left( T_{1 1}^2 \;  g(m_{\tilde{t}_1}, m_{\tilde{b}_1})
\label{eq:tbeps1}
\right.
\nonumber\\
 & + &  \left.
T_{1 2}^2 \; g(m_{\tilde{t}_2},m_{\tilde{b}_1}) -
T_{1 1}^2 \; T_{1 2}^2 \; g(m_{\tilde{t}_1}, m_{\tilde{t}_2}) \right),
\label{eq:eps1mst}
\end{eqnarray}
where $T_{ij}$ is the mixing matrix which diagonalizes the stop
mass matrix:
\begin{equation}
T {\cal{M}}_{st} T^{\dagger} =   {\cal{M}}^D_{st}.
\end{equation}
In  the above,
we have neglected the sbottom mixing angle, identifying
$\tilde{b}_L \equiv \tilde{b}_1$; this
is an excellent approximation for the low values of
$\tan\beta$ we are considering. The function $g(m_1,m_2)$
is directly related to the dependence of the variable $\epsilon_1$
on the top and bottom masses, Eq.(\ref{eq:tbeps1}),
\begin{equation}
g(m_1,m_2) = m_1^2 + m_2^2 - \frac{2 m_1^2 m_2^2}{ m_1^2 - m_2^2}
\ln \left( \frac{m_1^2}{m_2^2} \right).
\end{equation}

In the supersymmetric limit, $A_t = \mu = 0$, $\tan\beta = 1$,
the squark mixing vanishes, and, the weak eigenstates become
mass eigenstates with masses equal to their standard model
partners. It is easy to verify that the contribution to the
parameter $\epsilon_1$ of the stop--sbottom sector becomes
equal to the one of the top--bottom sector in this limit.
On the other hand, for small mixing and a soft supersymmetry
breaking parameter $m_Q^2 \gg m_t^2$,
\begin{equation}
\epsilon_1^{\tilde{t} - \tilde{b}} \simeq
\epsilon_1^{t - b}  \frac{m_t^2}{3 m_Q^2}.
\end{equation}
Hence, for sufficiently large values of the squark masses,
the squark contribution to the $\epsilon_1$ parameter vanishes.

The sleptons give a similar contribution to the parameter
$\epsilon_1$, although it is reduced by a factor 3 with
respect to Eq. (\ref{eq:eps1mst}), due to the color factor.
The only additional contribution that can become large is
the chargino one, if their masses are close to the
production threshold at the $Z^0$ peak, $m_{\tilde{\chi}_l}
\simeq M_Z/2$.
The derivative of the
chargino vacuum polarization contribution goes to large
values if the chargino masses approach the production threshold.
Indeed, it  behaves like
\begin{equation}
\Pi^{{'}} (M_Z^2) \simeq \left( M_Z^2 -  4 m_{\tilde{\chi}^+} \right)^{-1/2}.
\label{eq:threshold}
\end{equation}
The above expression formally diverges if charginos masses tend to
$M_Z/2$. However, Eq. (\ref{eq:e5}) losses its validity
when $m_{\tilde{\chi}^+} - M_Z/2 < \Gamma_Z$, what means that it can
only be trusted if the chargino masses are above  50 GeV \cite{BFC}.
In the following, we shall give here the dominant contribution to
$\Pi^{'}(M_Z^2)$ for sufficiently light
charginos. The
diagonalization of the chargino mass matrix is performed
by a bi--unitary tranformation
\begin{equation}
U^* \; {\cal{M}}_{ch} V^\dagger = {\cal{M}}_D.
\end{equation}
We can define the new matrices \cite{GS}
\begin{eqnarray}
U_{i j}^L & = &  \frac{1}{2} U^*_{i 2} U_{j 2}
- \cos^2\theta_W \delta_{i j}
\nonumber\\
U_{i j}^R & = &  \frac{1}{2} V_{i 2} V^*_{j 2}
- \cos^2\theta_W \delta_{i j}
\nonumber\\
X_{i j} & = &
U^L_{i j} U^{L *}_{i j} +
U^R_{i j} U^{R *}_{i j}
\nonumber\\
Y_{i j} & = &
U^L_{i j} U^{R *}_{i j} +
U^R_{i j} U^{L *}_{i j}
\label{eq:matrices}
\end{eqnarray}

Then,
the dominant (formally divergent in the limit $m_{\tilde{\chi}^+}
\rightarrow M_Z/2$)  chargino contributions to
the $\epsilon_1$ parameter are included in the definition of
the variable $e_5$, Eq. (\ref{eq:e5}), and are given by
\begin{eqnarray}
e_5 & = &
\frac{2 g_2^2}{\cos^2\theta_W} \sum_{i,j} \left\{
X_{i j} \left[
2 M_Z^2 \left( B^{'}_{21} (M_Z^2, M_i, m_j)
\right. \right. \right.
\nonumber\\
& - & \left. \left. \left.
B^{'}_1(M_Z^2, M_i, M_j) \right)
+
( M_j^2 - M_i^2) B_1^{'}(M_Z^2, M_i, M_j)
\right. \right.
\nonumber\\
& + & \left. \left.
M_i ( M_i X_{i j} - M_j Y_{i j} ) B_0^{'}(M_Z^2,M_i,M_j)
\right] \right\} ,
\end{eqnarray}
where $B_i^{'}$ simbolize the derivatives of the corresponding
Passarino--Veltman function \cite{PV}, which are given by
\begin{eqnarray}
B_0^{'}(M_Z^2,m_1^2,m_2^2) & = & \frac{1}{16 \pi^2}
\int_0^1 dx \; \frac{ x ( 1 - x )}{ \chi(m_1^2,m_2^2,x)}
\nonumber\\
B_{1}^{'}(M_Z^2,m_1^2,m_2^2) & = &
\frac{1}{ 16 \pi^2} \int_0^1 dx \;
\frac{ x^2 ( 1 - x)}{ \chi(m_1^2,m_2^2,x) }
\nonumber\\
B_{21}^{'}(M_Z^2,m_1^2,m_2^2) & = &
\frac{1}{16 \pi^2} \int_0^1 dx \;
\frac{ x^3 ( 1 - x )}{\chi(m_1^2,m_2^2,x)} ,
\end{eqnarray}
where
\begin{equation}
\chi(m_1^2,m_2^2,x) = m_1^2 + (m_2^2 - m_1^2 - M_Z^2) x + M_Z^2 \; x^2
\label{eq:chi}
\end{equation}
Observe that, for $m_1^2 = m_2^2$, the argument
\begin{equation}
\chi(m_1^2,m_1^2,x) = M_Z^2 \left[ (x - 1/2)^2 + (m_1^2/M_Z^2 - 1/4) \right],
\label{eq:chieq}
\end{equation}
and the derivative of the Passarino--Veltman functions listed
above become hence singular for $m_1^2 \rightarrow M_Z^2/4$.
\newpage
{}~\\
{\large{\bf {Appendix B.}}} \\

In this appendix, we include the relevant formulae for the
computation of the parameter $\epsilon_b$ in the
minimal supersymmetric standard model for the low
$\tan\beta$ regime. The main standard contribution come from the
Standard top quark - $W^+$ one loop diagram. This may be
expressed, within an excellent approximation for $M_t \geq
$ 160 GeV, as a  series in  the parameter
$r = M_t^2/M_W^2$, namely \cite{epsbst}

\begin{eqnarray}
 \epsilon_b^{SM} & = & - \frac{ \alpha }
{8 \pi \sin^2\theta_W}
\left[ r + 2.88 \log(r) - 6.716 +
\frac{\left( 8.368 \log(r) - 3.408 \right)}{r}
\right.
\nonumber\\
& + & \left.
\frac{ \left( 9.126 \log(r) + 2.26 \right) }{r^2} +
\frac{\left( 4.043 \log(r) + 7.41 \right)}{r^3} \right]
\end{eqnarray}

In the low $\tan\beta$ regime, the main contributions to the
$Z - b \bar{b}$ vertex,
associated with the
Higgs and supersymmetric particles come from the charged
Higgs contribution, which tends to enhance the Standard
Model signal, and the one coming from the chargino--stop
one loop contribution, which tends
to reduce the Standard Model signal.

The charged Higgs contribution is given by \cite{epsbs}
\begin{equation}
\epsilon_b^{H^+} = - \frac{ \alpha }{2 \pi \sin^2\theta_W}
F_b^{H^+}
\end{equation}
with
\begin{eqnarray}
F_b^{H^+} & = & \frac{M_t^2}{2 M_W^2 \tan^2\beta}
\left[ b_1(m_{H^+},M_t,M_b^2) v^{(t)}_L +
\left( \frac{M_Z^2}{\mu_R^2} c_6(m_{H^+},M_t,M_t)
\right. \right.
\nonumber\\
& - & \left. c_0(m_{H^+},M_t,M_t)
- \frac{1}{2}
\right) v^{(t)}_R
+ \frac{M_t^2}{\mu_R^2}\; c_2(m_{H^+},M_t,M_t) v^{(t)}_L
\nonumber\\
& + & \left.
c_0(M_t,m_{H^+},m_{H^+}) \left( \frac{1}{2} - \sin^2\theta_W
\right) \right] ,
\end{eqnarray}
where $\mu_R$ is a renormalization scale, $v_L^{(t)} = 0.5 -2
\sin^2\theta_w/3$ and $v_R^{(t)} = - 2 \sin^2\theta_W/3$, and
$b_1(a,b,c)$, $c_k(a,b,c)$ with $k = 0, 2, 6$ are the corresponding
reduced Passarino--Veltman functions. Since $m_b^2 \ll M_Z^2, M_t^2$,
they are well approximated by
\begin{eqnarray}
b_1(m_1,m_2,0) & = & \int_0^1 dx \;
x \log\left(
\frac{ m_1^2 x + m_2^2 (1 - x) }{ \mu_R^2 } \right)
\nonumber\\
c_0(m_1,m_2,m_3) & = & \int_0^1 dx \; \left(
\frac{  \tilde{\chi}(x)
\log\left[\tilde{\chi}(x)\right] - \tilde{\chi}(x) -
b(x) \log\left[b(x)\right]
+ b(x) }{a(x)} \right)
\nonumber\\
c_2(m_1,m_2,m_3) & = & \int_0^1 dx \;
\frac{ \log\left( \tilde{\chi}(x) \right)
- \log\left( b(x) \right) }{ a(x) }
\nonumber\\
c_6(m_1,m_2,m_3) & = & \int_0^1 dx \;
x \frac{ \log\left( \tilde{\chi}(x) \right)
- \log\left( b(x) \right) }{ a(x) },
\end{eqnarray}
and the arguments $a(x)$ and $b(x)$ are given by
\begin{eqnarray}
a(x) &  = & \frac{m_3^2 - m_1^2 - x M_Z^2}{\mu_R^2}
\nonumber\\
b(x) & = & \frac{m_1^2 + x \left( m_2^2 - m_1^2 \right)}
{\mu_R^2} ,
\end{eqnarray}
while
$\tilde{\chi}(x) = \chi(m_3^2,m_2^2,x)/\mu_R^2$ and
$\chi(m_3^2,m_2^2,x)$
has been defined in Eq.(\ref{eq:chi}).

The chargino contribution takes a somewhat more complicated
expression. It is given by \cite{epsbs}
\begin{equation}
\epsilon_b^{\tilde{\chi}^+} = - \frac{ \alpha }{2 \pi \sin^2\theta_W}
\left( F_b^{\tilde{\chi}^+}(M_t) - F_b^{\tilde{\chi}^+}(0) \right),
\end{equation}
where
\begin{equation}
F_b^{\tilde{\chi}^+}(M_t) =
F_b^{\tilde{\chi}^+ (a)}(M_t) +
F_b^{\tilde{\chi}^+ (b)}(M_t) +
F_b^{\tilde{\chi}^+ (c)}(M_t),
\end{equation}
and
\begin{eqnarray}
F_b^{\tilde{\chi}^+ (a)}(M_t) & = &
\sum_{i, j} b_1(  m_{\tilde{t},j},M_i,m_b^2)
\left| \Lambda_{j,i}^L \right|^2,
\nonumber\\
F_b^{\tilde{\chi}^+ (b)}(M_t) & = & \sum_{i,j,k} c_0(M_k, m_{\tilde{t},i},
 m_{\tilde{t},j}) \left( \frac{2}{3} \sin^2\theta_W \delta_{ij}
- \frac{1}{2} T^*_{i1} T_{j1} \right) \Lambda_{i k}^L
\Lambda^{* L}_{j k},
\nonumber\\
F_b^{\tilde{\chi}^+ (c)}(M_t) & = & \sum_{i,j,k}
\left\{ \left[ \frac{M_Z^2}{\mu_R^2} c_6( m_{\tilde{t},k},M_i,M_j)
- \frac{1}{2} - c_0( m_{\tilde{t},k},M_i,M_j) \right] U^R_{i j}
\right.
\nonumber\\
& + & \left.
\frac{M_i M_j}{\mu_R^2} c_2(  m_{\tilde{t},k},M_i,M_j ) U^L_{i j}
\right\} \Lambda^L_{k i} \Lambda^{* L}_{k j},
\end{eqnarray}
with
\begin{equation}
\Lambda^L_{i j} = T_{i 1} V^*_{j 1} - \frac{M_t}{\sqrt{2} M_W
\sin\beta} T_{i2} V^*_{j2}
\end{equation}
and $T_{ij}$ ($V_{ij}$, $U_{ij}$)
is the stop  (chargino)
mixing mass matrix (matrices) defined in Appendix A. Observe that
both the parameter $\Lambda^L_{i j}$ and the squark mass parameters
have a dependence on the top quark mass. Indeed, if the top quark mass
were negligible, the squark mass parameters would acquire an
approximately common value $m_{\tilde{t}}^2 \simeq m_0^2 + 7 M_{1/2}^2$.
The function $F_b^{\tilde{\chi}^+}(0)$ becomes, hence,
independent of the stop
mixing matrix (which is formally equal to the identity in the limit
$M_t = 0$).

\newpage
{}~\\
{\bf{FIGURE CAPTIONS}}\\
{}~\\
Fig. 1. Lightest stop mass as a function of the lightest chargino
mass, for the case of universal soft supersymmetry breaking parameters
at the grand unification scale and four different values of
the physical top quark mass $M_t = 160,\;165,\;175$ and 185 GeV.\\
{}~\\
Fig. 2. The same as figure 1, but for  the case I of non--universality
for the scalar mass parameters at $M_{GUT}$ :
$m_{H_1}^2(0) = 0$, $m_{H_2}^2(0) = 2 m_Q^2(0)$. \\
{}~\\
Fig. 3. The same as figure 1, but for the case II of non--universality
for the scalar mass parameters at $M_{GUT}$ :
$m_{H_1}^2(0) = 0$, $m_{H_2}^2(0) = 2 m_Q^2(0)$. \\
{}~\\
Fig. 4. Lightest CP--even Higgs mass as a function of the physical top
quark mass, for the values of $\tan\beta$, which for each value of $M_t$
corresponds to the top quark mass infrared fixed point solution
(crosses). Also shown in the figure is the upper bound on the Higgs
mass as a function of the top quark mass for values of $\tan\beta
\simeq 5$--10.\\
{}~\\
Fig. 5.  Lightest CP--even  Higgs mass as a function of the
lightest chargino mass for the case of universal scalar mass
parameters at $M_{GUT}$
and for  the same values of the physical
top quark mass $M_t = 165, 175$ and 185 GeV.\\
{}~\\
Fig. 6. The same as figure 5 but for the case I of non--universality
of the soft supersymmetry breaking parameters at $M_{GUT}$.\\
{}~\\
Fig. 7. The same as figure 4 but for the case II of non--universality
of the soft supersymmetry breaking parameters at $M_{GUT}$.     \\
{}~\\
Fig. 8. Dependence of the precision data variable  $\epsilon_1$
on the lightest chargino mass for the case of universal supersymmetry
breaking scalar mass parameters at $M_{GUT}$
and for three different
values of the top quark mass: $M_t = 165, 175$, 185 GeV. \\
{}~\\
Fig. 9. Dependence of the variables $\epsilon_1$, $\epsilon_b$
as a function of the lightest chargino mass
and the ratio of the
supersymmetric prediction for the branching ratio
$BR(b \rightarrow s \gamma)$ to the  standard model one,
as a function of the
supersymmetric mass parameter $\mu$,
for the case I of non--universality of the soft supersymmetry
breaking parameters at $M_{GUT}$ and a top quark mass $M_t =
175$ GeV. \\
{}~\\
Fig. 10. The same as Fig. 9 but for the case II of non--universality
of the soft supersymmetry breaking parameters at $M_{GUT}$.\\
{}~\\
Fig. 11. Dependence of the variable $\epsilon_b$ on the lightest
chargino mass for
the case of universal scalar mass  parameters at $M_{GUT}$
and for three different values of the top
quark mass: $M_t = 165, 175$ and 185 GeV. \\
{}~\\
Fig. 12. Dependence of the
ratio of the supersymmetric prediction for the
branching ratio $BR(b \rightarrow s \gamma)$
to the Standard Model one, as
a function of the supersymmetric mass parameter $\mu$ for the
case of universal scalar mass parameters at $M_{GUT}$ and three
different values of the top quark mass: $M_t = 165, 175$ and 185 GeV.\\
{}~\\
Fig. 13. Correlation between the variables $\epsilon_1$ and
$\epsilon_b$ for the case of universality of the soft
supersymmetry breaking parameters at $M_{GUT}$ and three different values
of the top quark mass: $M_t = 165, 175$ and 185 GeV. \\
{}~\\
Fig. 14. The same as Fig. 13, but for cases I and II of
non--universality of the scalar mass parameters at $M_{GUT}$
and a top quark mass $M_t = 175$ GeV.\\
{}~\\
Fig. 15. Correlation between the variables $\epsilon_b$ and
the ratio of the supersymmetric
prediction for the branching ratio
$BR(b \rightarrow s \gamma)$ to the standard model one, for the case of
universality of the soft supersymmetry breaking parameters at $M_{GUT}$
and three different values of the top quark mass:
$M_t = 165, 175$ and 185 GeV.\\
{}~\\
Fig. 16.  The same as Fig. 15, but for the cases I and II
of non--universality of the scalar mass parameters at $M_{GUT}$
and a top quark mass $M_t = 175$ GeV.\\

\end{document}